\documentclass{article}

\usepackage{microtype}
\usepackage{graphicx}
\usepackage{subcaption}
\usepackage{booktabs} % for professional tables

\usepackage{hyperref}

\usepackage[preprint]{icml2026}

\usepackage{amsmath}
\usepackage{amssymb}
\usepackage{mathtools}
\usepackage{amsthm}

\usepackage{multirow}
\usepackage{array}
\usepackage{ragged2e}
\newcolumntype{L}[1]{>{\RaggedRight\arraybackslash}p{#1}}
\newcolumntype{Y}{>{\RaggedRight\arraybackslash}X}
\usepackage{tabularx}

\usepackage[capitalize,noabbrev]{cleveref}

\theoremstyle{plain}
\newtheorem{theorem}{Theorem}[section]
\newtheorem{proposition}[theorem]{Proposition}
\newtheorem{lemma}[theorem]{Lemma}

\theoremstyle{definition}
\newtheorem{definition}[theorem]{Definition}

\theoremstyle{remark}

\usepackage[textsize=tiny]{todonotes}

\icmltitlerunning{Stabilizing Neural Emulation of Spatiotemporal Physical Dynamics through Denoising Koopman Latent Dynamics}

\begin{document}

\twocolumn[
  \icmltitle{Stabilizing Neural Emulation of Spatiotemporal Physical Dynamics \\ through Denoising Koopman Latent Dynamics}

  % It is OKAY to include author information, even for blind submissions: the
  % style file will automatically remove it for you unless you've provided
  % the [accepted] option to the icml2026 package.

  % List of affiliations: The first argument should be a (short) identifier you
  % will use later to specify author affiliations Academic affiliations
  % should list Department, University, City, Region, Country Industry
  % affiliations should list Company, City, Region, Country

  % You can specify symbols, otherwise they are numbered in order. Ideally, you
  % should not use this facility. Affiliations will be numbered in order of
  % appearance and this is the preferred way.

  % \icmlsetsymbol{cppd}{$\dagger$}
  % \icmlsetsymbol{equal}{$\dagger$}
  \icmlsetsymbol{equal}{$\star$}

  \begin{icmlauthorlist}
    \icmlauthor{Chenxu Wu}{equal,schin}
    \icmlauthor{Sixing Chen}{equal,schpy}
    \icmlauthor{Junming Hou}{equal,schin}
    \icmlauthor{Jiahan Huang}{schin}
    \icmlauthor{Jianqiao Zeng}{schai}%,acmai
    \icmlauthor{Peiming Luo}{schin}
    \icmlauthor{Liang Yan}{schma}
    \icmlauthor{Jian Wei You}{schin}
    % \icmlauthor{Firstname6 Lastname6}{sch,yyy,comp}
    % \icmlauthor{Firstname7 Lastname7}{comp}
    %\icmlauthor{}{sch}
    % \icmlauthor{Firstname8 Lastname8}{sch}
    % \icmlauthor{Firstname8 Lastname8}{yyy,comp}
    %\icmlauthor{}{sch}
    %\icmlauthor{}{sch}
  \end{icmlauthorlist}

  % \icmlaffiliation{labmm}{State Key Laboratory of Millimeter Waves, Southeast University, Nanjing, China}
  % \icmlaffiliation{schem}{Institute of Electromagnetic Space, Southeast University, Nanjing, China}
  \icmlaffiliation{schin}{School of Information Science and Engineering, Southeast University, Nanjing, China}
  \icmlaffiliation{schpy}{School of Physics and Engineering, Southeast University, Nanjing, China}
  \icmlaffiliation{schma}{School of Mathematics, Southeast University, Nanjing, China}
  % \icmlaffiliation{comp}{Company Name, Location, Country}
  \icmlaffiliation{schai}{AI$^3$ Institute, Fudan University}
  % \icmlaffiliation{acmai}{Shanghai Academy of AI for Science}

  \icmlcorrespondingauthor{Junming Hou}{junming\_hou@seu.edu.cn}
  \icmlcorrespondingauthor{Jian Wei You}{jvyou@seu.edu.cn}

  % You may provide any keywords that you find helpful for describing your
  % paper; these are used to populate the "keywords" metadata in the PDF but
  % will not be shown in the document
  \icmlkeywords{Machine Learning, ICML}

  \vskip 0.3in
]

% this must go after the closing bracket ] following \twocolumn[ ...

% This command actually creates the footnote in the first column listing the
% affiliations and the copyright notice. The command takes one argument, which
% is text to display at the start of the footnote. The \icmlEqualContribution
% command is standard text for equal contribution. Remove it (just {}) if you
% do not need this facility.

% Use ONE of the following lines. DO NOT remove the command.
% If you have no special notice, KEEP empty braces:
% \printAffiliationsAndNotice{}  % no special notice (required even if empty)
% Or, if applicable, use the standard equal contribution text:
\printAffiliationsAndNotice{\icmlEqualContribution}

\begin{abstract}
Koopman operator theory provides a principled framework for reduced-order modeling of nonlinear spatiotemporal dynamics through linear evolution in an observable space. In practice, however, learned finite-dimensional representations are typically approximately invariant, causing unresolved dynamics and accumulated closure errors. Recent approaches, inspired by the Mori–Zwanzig formalism, address these effects through deterministic memory or closure modeling, yet robust extrapolation beyond the training horizon remains challenging. Here, we propose KoopFlow, an efficient prediction-correction framework based on the key insight that finite-dimensional Koopman error accumulation can be reinterpreted as a progressive distributional degradation process.
Accordingly, KoopFlow treats Koopman predictions as noisy intermediate estimates, which are then transported toward the target latent distribution through a generative model, thereby mitigating accumulated errors while retaining computational efficiency and temporal flexibility.
% Accordingly, KoopFlow treats Koopman predictions as noisy intermediate estimates and leverages a pretrained flow matching model as a probabilistic transport bridge toward the target latent distribution, thereby mitigating accumulated errors while retaining computational efficiency and temporal flexibility.
Moreover, we reveal a non-monotonic relationship between Koopman spectral coherence and predictive accuracy. Experiments across diverse nonlinear dynamical systems demonstrate improved long-horizon prediction accuracy and stability over representative baselines, particularly beyond the training horizon.
\end{abstract}

\begin{figure}
\begin{center}
    % \vspace{-7mm}
    \includegraphics[width=\linewidth]{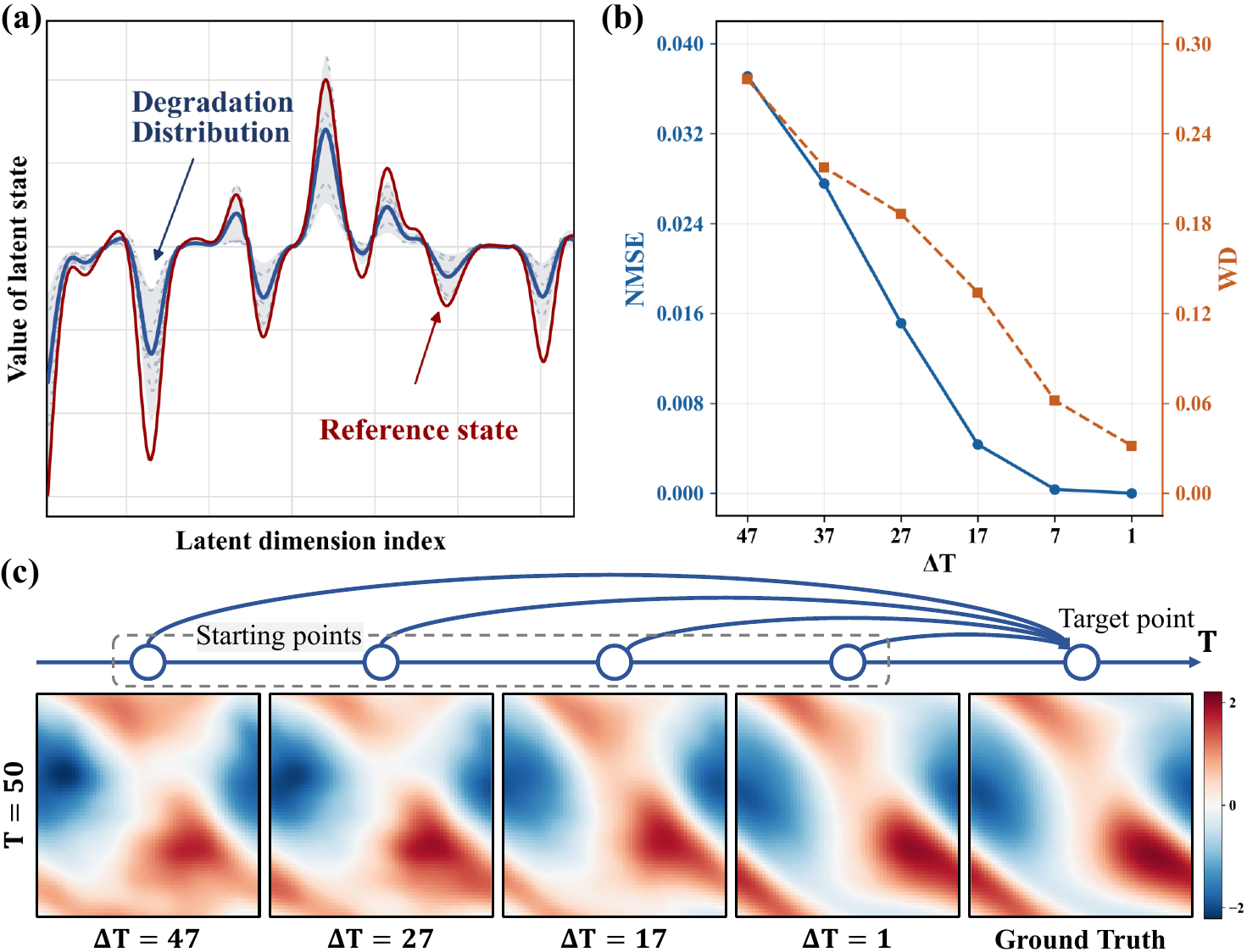}
    % \vspace{-0.8\baselineskip}
    % \vspace{-7mm}
    \caption{Motivation: (a) Reference latent state (red) and Koopman predictions (blue) evolved from different starting times along the same trajectory to the same target time. 
    % dashed curves and gray shading show individual predictions and their spread. 
    (b) NMSE and Wasserstein distance (WD) decrease as the prediction interval $\Delta T$ shortens. (c) Ground-truth and predicted fields at $t=50$ for different $\Delta T$.}
    \label{fig:motivation}
    % \vspace{-5mm}
\end{center}
\end{figure}

\section{Introduction}
Accurately predicting the spatiotemporal behavior of complex dynamical systems governed by time-dependent partial differential equations (PDEs) is central to scientific computing and engineering, with applications ranging from fluid dynamics~\cite{lagemann2026hydrogym} and climate forecasting~\cite{bi2023accurate} to electromagnetic wave propagation~\cite{cao2026physics}. Yet high-fidelity numerical simulations become increasingly costly as prediction horizons grow and spatial meshes are refined. Data-driven surrogates represented by neural operators~\citep{lu2021deeponet,li2021fno} offer a promising means of accelerating simulation after offline training.
Nevertheless, stable and accurate long-term prediction remains challenging, particularly when extrapolating beyond the training horizon. 
Moreover, generic operator learning architectures do not necessarily impose explicit dynamical structure on the learned representations, leaving their temporal evolution largely implicit.

\begin{figure*}[t]
\begin{center}
    \includegraphics[width=\linewidth]{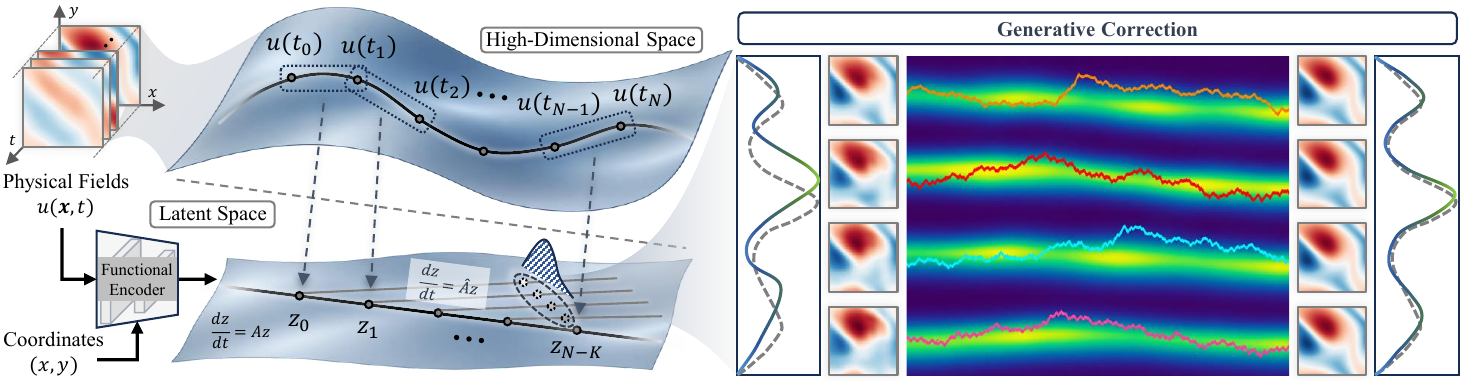}
    % \vspace{-7mm}
\end{center}
    \caption{Overview of our proposed KoopFlow. High-dimensional field snapshots are embedded into a latent space via a functional encoder and evolved by a Koopman operator, where finite-dimensional approximation leads to degraded predictions (Left). A lightweight generative model transports these states toward the target latent distribution, yielding high-fidelity predictions (Right).}
    \label{fig:framework}
    % \vspace{-3mm}
\end{figure*}

In contrast, Koopman operator theory provides a principled framework for modeling dynamical systems by representing nonlinear state evolution through the linear evolution of observables in an appropriate function space~\citep{koopman1931hamiltonian,mezic2005spectral,brunton2021modern}. Building on this principle, learning-based Koopman operator frameworks have attracted increasing attention for reduced-order modeling of spatiotemporal dynamics.
Despite this appeal, the practical realization is constrained by the gap between the infinite-dimensional Koopman representation and its finite-dimensional approximations.
For general nonlinear systems, the existence of a finite-dimensional Koopman-invariant subspace capturing state dynamics is not guaranteed, whereas practical learning-based models must operate with a finite number of learned observables~\citep{lusch2018deep,xiong2024kno}.
Recent work addresses the quality of these learned representations through information-theoretic regularization, balancing latent linear predictability and expressiveness while mitigating mode collapse~\citep{cheng2026information}.
Nevertheless, the resulting observable subspace need not be invariant, leaving part of the dynamics unresolved~\citep{brunton2016koopman,brunton2021modern}.
From the Mori-Zwanzig perspective, restricting high-dimensional dynamics to such a non-closed resolved subspace gives rise to memory and unresolved forcing terms beyond the Markovian evolution of the resolved variables~\citep{chorin2000optimal,lin2021mori}.
Most existing deep Koopman models, however, assume a closed Markovian latent evolution and therefore do not account for these memory and forcing terms. The resulting closure errors can accumulate or be amplified during recursive rollouts, degrading long-term prediction stability and accuracy. Recent studies have begun to incorporate explicit memory or closure mechanisms into Koopman models~\citep{gupta2025mzae,lu2026merlin}.
Yet it remains challenging to achieve computationally tractable closures that model the influence of these unresolved effects.
In this work, rather than explicitly learning deterministic approximations of unresolved effects, we model their aggregate influence on finite-dimensional Koopman predictions statistically. For exact deterministic dynamics, the semigroup property ensures that predictions from different earlier states along the same trajectory agree at a fixed target~\cite{mezic2005spectral}. However, finite-dimensional approximations~\cite{mauroy2020koopman} may violate this consistency, producing horizon-dependent deviations (Figure~\ref{fig:motivation}). We term the empirical distribution of these predictions the \emph{Koopman degradation distribution}, capturing the combined effects of unresolved dynamics and approximation errors on predicted latent states. From this perspective, long-horizon Koopman prediction can be formulated as a distribution-correction problem, with the goal of reducing the mismatch between predicted distributions and reference latent states while preserving trajectory-specific information.
This insight gives rise to KoopFlow, a prediction–correction framework (Figure~\ref{fig:framework}) that combines efficient Koopman evolution with generative correction. 
Specifically, (i) we first learn a finite-dimensional Koopman operator over an empirically selected temporal horizon, motivated by our finding that effective Koopman learning requires balancing long-term information preservation with transient dynamics fidelity; and (ii) the resulting prediction is then anchored at one endpoint of a stochastic process and denoised by a generative correction model to recover the target latent state.
% the resulting prediction is mapped to an intermediate noisy latent state and subsequently denoised by a pretrained flow-matching model.
This mechanism mitigates accumulated approximation errors and improves long-horizon predictive stability and accuracy.
The main contributions of this paper are summarized as follows:
\begin{itemize}
\item We reinterpret finite-dimensional Koopman error accumulation as a progressive distributional degradation process and propose KoopFlow, an efficient prediction-correction framework that couples structured linear Koopman evolution with generative correction for stable and flexible long-horizon forecasting.
\item We identify a non-monotonic relationship between Koopman spectral coherence and predictive accuracy, showing that effective Koopman learning requires balancing long-term information preservation with transient dynamics fidelity.
\item Extensive experiments across diverse nonlinear dynamical systems demonstrate that KoopFlow consistently improves long-horizon prediction accuracy and stability over representative baselines, particularly beyond the training horizon, while maintaining a favorable accuracy-efficiency trade-off.
\end{itemize}

\section{Related work}
\textbf{Koopman Learning.} Koopman learning has emerged as an effective paradigm for modeling nonlinear dynamical systems by learning observable representations with approximately linear evolution~\citep{takeishi2017learning,lusch2018deep,azencot2020forecasting}. However, finite-dimensional Koopman representations inevitably discard unresolved dynamics, which appear as memory and forcing terms under the Mori-Zwanzig formalism~\citep{mori1965transport,zwanzig1973nonlinear}. Existing methods mainly address these limitations by improving Koopman representations~\citep{haseli2021learning,cheng2026information} or introducing memory and closure corrections~\citep{gupta2025mzae,lu2026merlin}. 
The former reduces projection error but neglects the intrinsic dynamical structure of the underlying physical system, while the latter better captures unresolved dynamics at the cost of additional states or corrective dynamics. Both primarily focus on improving forward evolution. In contrast, we adopt a trajectory consistency view and show that accumulated finite-dimensional Koopman errors induce a progressive, horizon-dependent degradation distribution, shifting the focus from error suppression during propagation to correct the degraded latent dynamics.

\textbf{Generative Modeling.}
Generative models, particularly diffusion and flow matching models~\cite{shysheya2024conditional}, have recently attracted increasing attention for learning complex PDE dynamics~\citep{huang2024diffusionpde,liu2025diff,hou2026cfo}, producing statistically plausible predictions with improved long-horizon stability over deterministic forecasts. However, generative modeling in the high-dimensional physical state space remains computationally expensive, especially at high spatial resolutions~\citep{huang2024diffusionpde,kohl2026benchmarking,hou2026cfo,holzschuh2026p3d}.
Recent studies therefore perform generative modeling in compressed latent spaces~\citep{du2024conditional,zhou2025text2pde,rozet2025lola}. Despite improved efficiency, their performance remains sensitive to representation quality, while latent compression may compromise fine-scale fidelity or dynamical consistency.
Moreover, bridge-based generative methods~\cite{de2021diffusion,wang2021deep} have attracted increasing attention in image processing~\cite{li2023bbdm,liu20232,yue2023image,lan2025schrodinger} because they explicitly model stochastic transitions between source and target distributions, providing a natural framework for conditional distribution translation while preserving source-specific information. This paradigm is also beginning to be explored for PDE modeling and solving~\cite{bu2026sparsely}.
In this context, rather than directly learning the full dynamical evolution with a generative model, we anchor structured Koopman predictions and apply probabilistic correction to recover the target states.
% Rather than using generative models to directly learn the dynamical evolution, we employ a trained flow matching model to probabilistically correct structured Koopman predictions. 
This design couples efficient linear Koopman propagation with generative correction, balancing accuracy and computational efficiency.

\section{Methodology}
Our methodology proceeds as follows: (1) the problem formulation (Section~\ref{sec:problem}) and continuous-time Koopman learning (Section~\ref{sec:cont-koop}); (2) a trajectory-consistency analysis of finite-dimensional Koopman dynamics, which motivates a horizon-dependent degradation distribution (Section~\ref{sec:koop_degra}); 
% (3) a theoretical guarantee for generative recovery from the degraded distribution to the target latent state (section~\ref{sec:gen_corr});
(3) the construction of a continuous stochastic bridge connecting reference and degraded latent states, providing the motivation for generative correction (Section~\ref{sec:gen_corr});
(4) the KoopFlow framework built upon these insights (Section~\ref{sec:koop_frame}).

\subsection{Problem Formulation}
\label{sec:problem}
We consider a parameterized spatiotemporal dynamical system governed by time-dependent partial differential equations (PDEs). Let $\Omega\subset\mathbb{R}^d$ be the spatial domain and $u(t,x)\in\mathbb{R}^c$ denote the physical state at time $t$ and location $x\in\Omega$. The governing equations are expressed as:
\begin{equation}
\label{eq:pde_system}
\begin{aligned}
    \partial_t u(t,x)
    =\mathcal{F}[u(t);\mu](x)&,
    && x\in\Omega,\ t>0,\\
    u(0,x)=u_0(x)&,
    && x\in\Omega,\\
    \mathcal{B}[u(t);\mu](x)=0&,
    && x\in\partial\Omega,\ t>0.
\end{aligned}
\end{equation}
Here, $\mathcal{F}$ denotes the governing differential operator, $\mu$ represents the physical parameters, $\mathcal{B}$ specifies the boundary conditions, and $u_0$ is the initial state.
We assume access to discrete trajectory observations $u_n=u(n\Delta t)$ sampled at temporal intervals $\Delta t$. Given the most recent $K$ observations $u_{n-K+1:n}$, together with the
physical parameters $\mu$ and boundary conditions $\mathcal{B}$, our objective is to predict the future physical state $u_{n+h}$ at a forecasting horizon $h\Delta t$, where
$h\in\mathbb{N}_{>0}$. To efficiently model the underlying dynamics and enable flexible temporal prediction, we introduce a continuous-time Koopman formulation.

\subsection{Continuous-Time Koopman Learning}
\label{sec:cont-koop}
We learn a latent Koopman representation in which nonlinear spatiotemporal dynamics are approximated by continuous-time linear evolution, enabling flexible prediction horizons. A functional encoder $E_\theta$ maps historical observations, potentially sampled at irregular spatial locations and varying resolutions, to a fixed-dimensional latent state $z_n=E_\theta(u_{n-K+1:n},\mu,\mathcal{B})\in\mathbb{R}^m$.
A paired functional decoder $D_\theta$ reconstructs the physical field from the latent state.
Based on the Koopman operator theory, we approximate the latent dynamics using a learned finite-dimensional linear generator $A\in\mathbb{R}^{m\times m}$, which is defined as:
\begin{equation}
\frac{\mathrm{d}z}{\mathrm{d}t}=Az.
\end{equation}
The learned latent evolution operator is given by:
\begin{equation}
\mathcal{K}_A^{\tau}
=\exp(\tau A),
\quad \tau\geq0.
\end{equation}
This operator satisfies the semigroup property $\mathcal{K}_A^{\tau_1+\tau_2} =\mathcal{K}_A^{\tau_1}\mathcal{K}_A^{\tau_2}$ and allows direct propagation to the prescribed lead time~\citep{mezic2005spectral}.
Accordingly, for a forecasting horizon $h\Delta t$, the predicted latent state and the corresponding physical field are computed as:
\begin{equation}
\label{eq:koopman_forecast}
\begin{aligned}
    \widetilde{z}_{n+h}
    &=\mathcal{K}_A^{h\Delta t}z_n
    =\exp(h\Delta t A)z_n,\\
    \widetilde{u}_{n+h}
    &=D_\theta(\widetilde{z}_{n+h}).
\end{aligned}
\end{equation}

\subsection{Koopman Evolution as Progressive Degradation}
\label{sec:koop_degra}
Although continuous-time deep Koopman models enable direct prediction at prescribed lead times, finite-dimensional approximations generally incur closure and estimation errors. The Mori-Zwanzig formalism describes the influence of unresolved dynamics through memory and orthogonal-dynamics terms~\citep{mori1965transport}. Here, we examine how these approximation errors propagate and induce trajectory inconsistency among predictions initialized differently along the same latent trajectory.
% By combining error accumulation analysis with the consistency of deterministic dynamics, we characterize the resulting degradation distribution, motivating the subsequent generative correction.
This analysis introduces a horizon-dependent degradation distribution and thus motivates the subsequent generative correction.

% Let $\psi_{0:T}=\{\psi_0,\psi_1,\ldots,\psi_T\}$ denote a reference infinite-dimensional trajectory obtained by encoding the observed dynamics and $\mathcal K:\mathcal H\to\mathcal H$ denote the exact Koopman operator on the infinite-dimensional observable space. For a target future state $\psi_K$, all historical states along the same trajectory should evolve to an identical state under the exact Koopman operator:

We first consider an exact feature representation of the underlying dynamics.
Let $\mathcal H$ be an infinite-dimensional Hilbert space, and let $\psi_{0:T}=\{\psi_0,\psi_1,\ldots,\psi_T\}$ denote a reference feature trajectory.
Assume that a bounded linear evolution operator $\mathcal K:\mathcal H\to\mathcal H$
satisfies $\psi_{n+1}=\mathcal K\psi_n$.
% This exact feature representation is introduced for theoretical analysis and does not assume exact linear closure of the learned finite-dimensional latent variables.
For a target future state $\psi_K$, the exact evolution satisfies:
\begin{equation}
    \psi_K=\mathcal{K}^{K-i}\psi_i,\quad i=0,\ldots,K-1.
\end{equation}
However, practical Koopman models rely on finite-dimensional approximations.
% of the infinite-dimensional Koopman operator. 
% Let $\widehat{\mathcal K}$ denote the learned Koopman operator and $\mathcal K$ the exact Koopman operator. 
% Then with the mismatch modeled as the operator perturbation $\Delta\mathcal K=\widehat{\mathcal K}-\mathcal K$, we analyze the accumulation of Koopman approximation errors.
Let $\widehat{\mathcal K}\in\mathbb R^{m\times m}$ denote the learned finite-dimensional Koopman operator. We introduce a projection operator $\mathcal P_m:\mathcal H\to\mathbb R^m$ and a lifting operator $\mathcal I_m:\mathbb R^m\to\mathcal H$ to map between the infinite-dimensional observable space and the finite-dimensional latent space, which satisfy the projection-lifting consistency condition $\mathcal P_m\mathcal I_m=I_m$. The learned operator is then lifted to $\mathcal H$ as $\widehat{\mathcal K}_{\mathcal H}=\mathcal I_m\widehat{\mathcal K}\mathcal P_m$ and the ideal finite-dimensional operator is $\mathcal K^\star=\mathcal P_m\mathcal K\mathcal I_m$. Accordingly, we define the total operator perturbation as $\Delta\mathcal K=\widehat{\mathcal K}_{\mathcal H}-\mathcal K=\mathcal I_m\widehat{\mathcal K}\mathcal P_m-\mathcal K$ and analyze the accumulation of Koopman approximation errors. In this setting, $\Delta\mathcal K$ admits the following exact decompositions:
\begin{equation}
    \Delta\mathcal K=\underbrace{\mathcal I_m(\widehat{\mathcal K}-\mathcal K^\star)\mathcal P_m}_{\Delta\mathcal K_{\mathrm{learn}}}
+(\underbrace{\mathcal I_m\mathcal K^\star\mathcal P_m-\mathcal K}_{\Delta\mathcal K_{\mathrm{red}}}),
\end{equation}
% which indicates that the error of the learned Koopman process can be decomposed into two components: one is the deviation between the learned and ideal finite-dimensional operator while the other is the truncation error from infinite dimensions. 
which indicates that the error of the learned Koopman operator can be decomposed into two distinct components: one is the deviation between the learned and ideal finite-dimensional operators, while the other is the truncation error introduced by finite-dimensional reduction. Based on this, we further analyze the accumulation of the error.
% Then we have the following lemma.
\begin{lemma}[Accumulation of Koopman Approximation Error]
\label{lem:err_accumulation}
    For any historical state $\psi_i\in\mathcal H$, define the $n$-step rollout deviation under the perturbed Koopman operator as $e_n^{(i)}=\widehat{\mathcal K}_{\mathcal H}^{n}\psi_i-\mathcal K^{n}\psi_i$. 
    % When $\|\Delta\mathcal K\|$ is sufficiently small, the accumulated error can be approximated as
    As $\|\Delta\mathcal K\|\to 0$, the accumulated error can be expressed as: 
    \begin{equation}
    e_n^{(i)}=\sum_{j=0}^{n-1}\mathcal K^{n-1-j}
    \Delta\mathcal K\mathcal K^j \psi_i+\mathcal O(\|\Delta\mathcal K\|^2\|\psi_i\|),
    \end{equation}
\end{lemma}
The proof is provided in Appendix~\ref{app:proof_lamma1}. Lemma~\ref{lem:err_accumulation} establishes that finite-dimensional Koopman errors accumulate under repeated evolution and vary with the propagation horizon. Following this, with $z_i=\mathcal P_m\psi_i$ and $e_{n,\mathrm{fd}}^{(i)}=\widehat{\mathcal K}^n z_i-z_{i+n}$, the prediction error in finite-dimension space can be obtained from $e_{n,\mathrm{fd}}^{(i)}=\mathcal P_m e_n^{(i)}$, where $z_{i+n}=\mathcal P_m\psi_{i+n}=\mathcal P_m\mathcal K^{n}\psi_i$. The operator perturbation formulation can also be interpreted as a continuous-time closure error, where the discrepancy between the true dynamics and the finite-dimensional Koopman representation accumulates along the latent trajectory. We next characterize this horizon-dependent growth.

\begin{proposition}[Horizon-dependent Koopman Rollout Error]
\label{prop:koop_roll_err}
    Assume that the exact Koopman evolution operator is bounded as $\|\mathcal{K}\|\leq\rho$, and that the lifted finite-dimensional approximation satisfies $\|\Delta\mathcal{K}\|\leq\epsilon$.
    % , where $\Delta\mathcal K=\widehat{\mathcal K}_{\mathcal H}-\mathcal K$. 
    % The $n$-step rollout deviation introduced by the finite-dimensional Koopman approximation satisfies:
    For any reference state $\psi_i$ and integer $n\geq1$, the rollout error $e_n^{(i)}= \widehat{\mathcal K}_{\mathcal H}^{\,n} \psi_i-\mathcal K^n\psi_i$ satisfies:
    % \begin{equation}
    %     \|e_n^{(i)}\|\leq n\rho^{n-1}\epsilon\|\psi_i\|+\mathcal{O}(\epsilon^2).
    % \end{equation}
    \begin{equation}
    \|e_n^{(i)}\|\leq\left[(\rho+\epsilon)^n-\rho^n\right]\|\psi_i\|.
    \end{equation}
    In particular, for fixed $n$ and $\epsilon\to0$,
    \begin{equation}
    \|e_n^{(i)}\|\leq\left[n\rho^{n-1}\epsilon+\mathcal O_n(\epsilon^2)\right]\|\psi_i\|.
    \end{equation}
\end{proposition}
The proof is provided in Appendix~\ref{app:proof_prop1}. 
% Let $z_i=\mathcal P_m\psi_i$ and define the finite-dimensional rollout error as
% \begin{equation}
% e_{n,\mathrm{fd}}^{(i)}=\widehat{\mathcal K}^{\,n}z_i-z_{i+n}.
% \end{equation}
% Under the projection-lifting consistency condition $\mathcal P_m\mathcal I_m=I_m$,
% we have $e_{n,\mathrm{fd}}^{(i)}=\mathcal P_m e_n^{(i)}$.
Further assuming $\|\mathcal P_m\|\leq1$, the finite-dimensional rollout error satisfies:
\begin{equation}
\|e_{n,\mathrm{fd}}^{(i)}\|\leq\left[n\rho^{n-1}\epsilon+\mathcal O_n(\epsilon^2)\right]\|\psi_i\|.
\end{equation}
This result characterizes the horizon-dependent accumulation of Koopman approximation errors. The bound indicates that repeated propagation can amplify approximation errors, particularly when $\rho>1$. 
% although monotonic growth of the actual prediction error is not guaranteed.

% Further assuming $\|\mathcal P_m\|\leq 1$, we then have $e_{n,\mathrm{fd}}^{(i)}\leq n\rho^{n-1}\epsilon\|\psi_i\|+\mathcal{O}(\epsilon^2)$. This result characterizes the horizon-dependent accumulation of Koopman approximation error, showing that when the bound of operator is more than 1, longer rollouts may introduce larger deviation due to repeated error propagation.

Motivated by this horizon-dependent error accumulation, we interpret finite-dimensional Koopman predictions from the perspective of trajectory consistency. \textit{For deterministic dynamics, historical states along the same trajectory must evolve to the same reference state at a fixed target time~\cite{nayak2025temporally,shikhman2026semigroup}}. 
Given a target state $z_K$, the approximate Koopman predictions satisfy:
% exact and 
% \begin{equation}
% \begin{aligned}
%     z_K&={\mathcal K^*}^{K-i}z_i,
%     \quad \\
%     \widehat z_K^{(i)}
%     :&=\widehat{\mathcal K}^{K-i}z_i
%     =z_K+e_{K-i,\mathrm{fd}}^{(i)},
%     \ i=0,\ldots,K-1.
% \end{aligned}
% \end{equation}
\begin{equation}
    \widehat z_K^{(i)}
    :=\widehat{\mathcal K}^{K-i}z_i
    =z_K+e_{K-i,\mathrm{fd}}^{(i)},
    \ i=0,\ldots,K-1.
\end{equation}
Therefore, finite-dimensional Koopman approximations may introduce trajectory inconsistency, as predictions initialized from different historical states can deviate from the same reference state. We regard $\{\widehat z_K^{(i)}\}_{i=0}^{K-1}$ as degraded predictions of $z_K$, which naturally induces the following empirical distribution.

\begin{definition}[Koopman-induced Empirical Degradation Distribution]
\label{def:koop_dist}
    Consider a reference latent trajectory
    $z_{0:K}=\{z_0,z_1,\ldots,z_K\}$ and a historical state index
    $I\sim p(I)$ supported on $\{0,\ldots,K-1\}$.
    The degraded prediction is defined as $Z_K^{\mathrm{deg}}=\widehat{\mathcal{K}}^{K-I}z_I$.
    Then the resulting empirical degradation distribution induced by finite-dimensional Koopman prediction is given by:
    \begin{equation}
        q_{\mathrm{deg}}(\widetilde z_K
        \mid z_{0:K})
        =
        \sum_{i=0}^{K-1}p(I=i)
        \delta_{\widehat{\mathcal K}^{K-i}z_i}(\widetilde z_K),
    \end{equation}
    where $\delta_a$ denotes the Dirac probability measure concentrated at $a$.
    This distribution is induced by predictions initialized from different historical states without externally injected perturbations.
    % To characterize the deviations of degraded predictions, under the definition of $e_{n,\mathrm{fd}}^{(i)}$, the corresponding empirical bias and covariance are given by:
    % \begin{equation}
    % \label{eq:degradation_moments}
    % \begin{aligned}
    % b_K&=\sum_{i=0}^{K-1}p(I=i)e_{K,\mathrm{fd}}^{(i)},\\
    % c_K&=\sum_{i=0}^{K-1}p(I=i)
    % (e_{K,\mathrm{fd}}^{(i)}-b_K)(e_{K,\mathrm{fd}}^{(i)}-b_K)^\top.
    % \end{aligned}
    % \end{equation}
    % These quantities characterize the systematic deviation and
    % the variability of Koopman-degraded predictions across
    % historical starting points.
\end{definition}
This probabilistic formulation provides a statistical characterization of horizon-dependent Koopman prediction errors. The resulting distribution captures the variability among predictions initialized from different historical states and propagated to the same target time, reflecting trajectory inconsistency induced by finite-dimensional Koopman approximations.
Accordingly, we formulate long-horizon forecasting as a conditional latent state correction problem, aiming to mitigate accumulated prediction errors while preserving trajectory-specific information.
%% prop:koop_degra --------------------------------------------

\subsection{Generative Correction}
\label{sec:gen_corr}
\begin{figure}[t]
    \begin{center}
    % \vspace{-3mm}
    \includegraphics[width=0.9\linewidth]{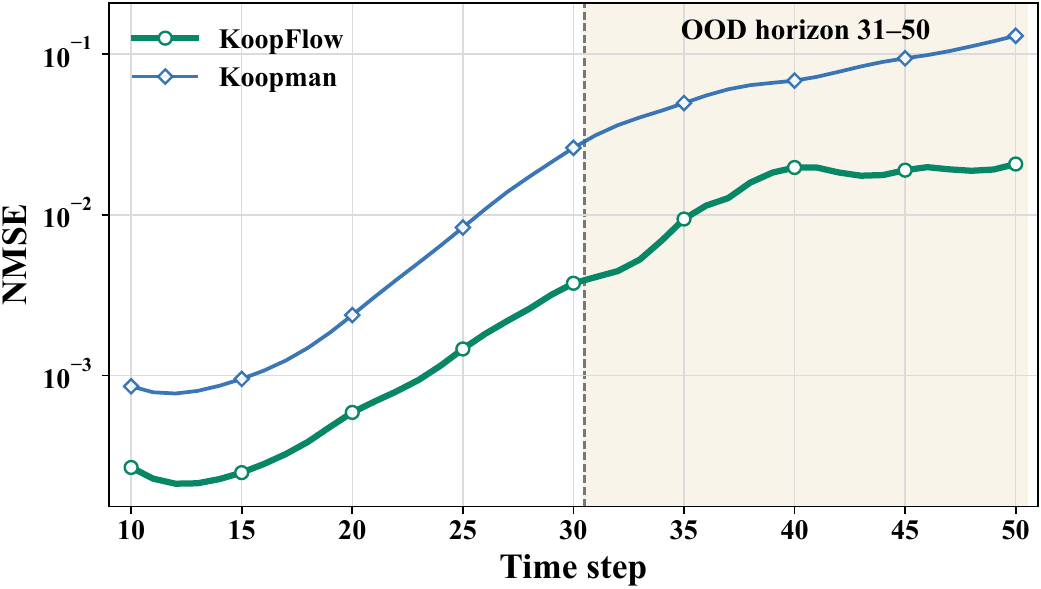}
    % \vspace{-0.8\baselineskip}
    % \vspace{-9mm}
\end{center}
    \caption{Long-horizon error accumulation for Koopman prediction with and without generative correction.}
    \label{fig:flow_correct}
    % \vspace{-5mm}
\end{figure}

Building on the Koopman-induced degradation distribution, our goal is to correct a degraded prediction $\widetilde z_K$ toward its reference latent state $z_K$ while preserving the trajectory-specific information carried by Koopman evolution. 
We approach this distribution correction using a trained generative prior over reference latent states. Rather than assuming that Koopman approximation errors follow the forward process of a generative model, we anchor the predicted state as the noisy endpoint of the stochastic process and then exploit the learned generative dynamics to correct it. This decouples the prediction degradation induced by approximate Koopman evolution from the auxiliary generative path used for correction.
% We approach this distribution correction using a pretrained generative prior over reference latent states. Rather than assuming that Koopman approximation errors follow the forward process of a generative model, we map the predicted state to an intermediate noisy representation and then exploit the learned generative dynamics to correct it. This decouples the prediction degradation induced by approximate Koopman evolution from the auxiliary generative path used for correction.
Let $\{Z_s\}_{s\in[0,1]}$ denote a continuous path connecting the reference latent state and the degraded prediction, with $Z_0=z_K$ and $Z_1=\widetilde z_K$. 
Figure~\ref{fig:flow_correct} shows that generative correction (i.e., KoopFlow) substantially reduces long-horizon error accumulation relative to Koopman prediction, with increasingly pronounced improvements beyond the training horizon.
\begin{theorem}[Existence of Koopman-induced Reverse Process]
\label{them:ext_reverse}
    Given the Koopman-induced degradation distribution $q_{\mathrm{deg}}(\widetilde z_K|z_{0:K})$ in Definition~\ref{def:koop_dist}, there exists a continuous stochastic process $\{Z_s\}_{s\in[0,1]}$ satisfying:
    \begin{equation}
        Z_0=z_K,\qquad Z_1=\widetilde z_K,
        \quad\widetilde z_K\sim q_{\mathrm{deg}}(\cdot|z_{0:K}).
    \end{equation}
    Moreover, its time-reversed process $\bar Z_s:=Z_{1-s}$ defines a continuous reverse process from the degraded state to the target latent state, satisfying $\bar Z_0=\widetilde z_K$ and $\bar Z_1=z_K$.
\end{theorem}
% \begin{theorem}[Existence of a Koopman-induced Continuous Stochastic Path]
% \label{them:ext_reverse}
%     Given a target latent state $z_K$ and the Koopman-induced degradation distribution
%     $q_{\mathrm{deg}}(\widetilde z_K|z_K)$ in Definition~\ref{def:koop_dist},
%     let $\widetilde z_K\sim q_{\mathrm{deg}}(\cdot|z_K)$.
%     A continuous stochastic process $\{Z_s\}_{s\in[0,1]}$ connecting
%     the target and degraded latent states can be constructed as:
%     \begin{equation}
%         Z_s=(1-s)z_K+s\widetilde z_K,
%         \quad s\in[0,1],
%         \quad s.t.\ Z_0=z_K,\quad Z_1=\widetilde z_K.
%     \end{equation}
%     Moreover, its time-reversed parameterization
%     $\bar Z_s:=Z_{1-s}$ defines a continuous path from the degraded latent state to the target latent state, satisfying $\bar Z_0=\widetilde z_K,\qquad \bar Z_1=z_K$.
% \end{theorem}
The proof is provided in Appendix~\ref{app:proof_theo1}. Theorem~\ref{them:ext_reverse} establishes the existence of a reverse stochastic process for recovering target latent states from Koopman-induced degradation, motivating the generative correction:
% introduced below:
\begin{equation}
    q_{\mathrm{deg}}(\widetilde z_K|z_K)
    \rightarrow p_\phi(z_K|\widetilde z_K).
\end{equation}
% This process demonstrates that Koopman evolution can provide an efficient prediction prior and the generative model then performs distribution-level correction. 
This process demonstrates that Koopman evolution provides an efficient prediction prior, upon which the generative model subsequently performs distribution-level correction to mitigate accumulated approximation errors.
% In the following, we further characterize this process under the Gaussian degradation model derived in Proposition~\ref{prop:koop_degra}.
% \begin{remark}[Gaussian Form]
%     Under the Gaussian degradation in Proposition~\ref{prop:koop_degra}, the stochastic process in Theorem~\ref{them:ext_reverse} can be interpreted as a continuous uncertainty accumulation process. Specifically, the discrete covariance sequence $\{\Sigma_n\}$ can be embedded into a monotonic schedule $\Sigma(s)$ with $\Sigma(s_1)\preceq\Sigma(s_2)$ for $s_1<s_2$, leading to a stochastic trajectory $Z_s=z_K+\epsilon_s$, where $\epsilon_s\sim\mathcal{N}(0,\Sigma(s))$. The boundary conditions satisfy $Z_0=z_K$ and $Z_1\sim q_{\mathrm{deg}}(\cdot|z_K)$.
% \end{remark}
% This interpretation reveals that the Koopman-induced degradation process shares a similar structure with diffusion-like stochastic processes, where uncertainty gradually increases from accurate latent states to degraded predictions.
Compared with Koopman-only prediction, Figure~\ref{fig:flow_correct} demonstrates that generative correction reduces prediction error by nearly an order of magnitude and exhibits improved stability at the terminal horizon.
\subsection{Framework of KoopFlow}
\label{sec:koop_frame}
Building on the above statistical insight, we develop KoopFlow, a prediction–correction framework that is implemented through a two-stage pipeline.

\begin{figure}[t]
    \begin{center}
    % \vspace{-7mm}
    \includegraphics[width=0.95\linewidth]{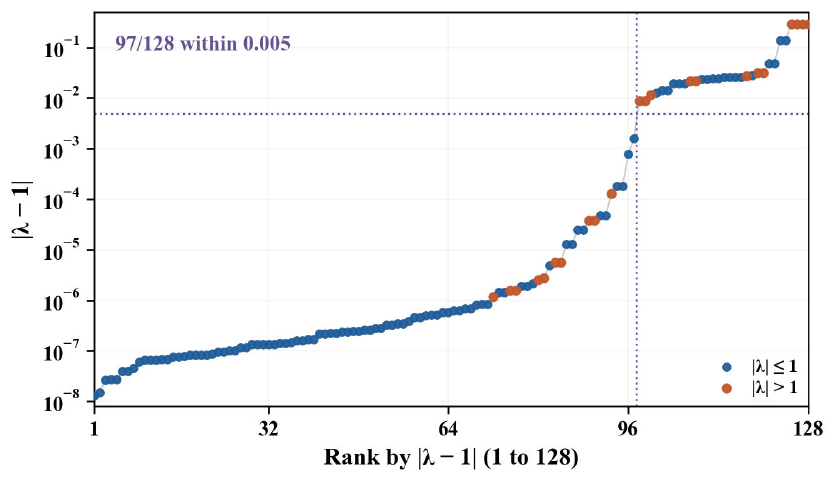}
    % \vspace{-0.8\baselineskip}
    % \vspace{-9mm}
\end{center}
    \caption{Koopman eigenvalue deviations from 1 at the empirically selected training horizon.}
    \label{fig:eigenvalue}
    % \vspace{-7mm}
\end{figure}

\textbf{Training of Koopman Dynamics.}
We first jointly train the encoder $E_\theta$, decoder $D_\theta$, and finite-dimensional Koopman generator $A$.
Given $z_n=E_\theta(u_{n-K+1:n},\mu,\mathcal{B})$, we obtain the $h$-step Koopman prediction $\widetilde z_{n,h}=\exp(h\Delta t A)z_n$. 
The optimization objective $\mathcal{L}_{\mathrm{Stage\,I}}$ is defined as follows:
\begin{equation}
\label{eq:koopman_objective}
\begin{aligned}
    &\mathcal{L}_{\mathrm{Stage\,I}}
    = \mathbb{E}_{n}\Big[
    \lambda_{\mathrm{rec}}\|D_{\theta}(z_n)-u_n\|_2^2 \\
    &+ \frac{1}{L}\sum_{h=1}^{L}
    \Big(\lambda_{\mathrm{lat}}\|\widetilde z_{n,h}-z_{n+h}\|_2^2 \\
    &+\lambda_{\mathrm{pred}}\|D_{\theta}(\widetilde z_{n,h})-u_{n+h}\|_2^2
    \Big)
    \Big],
\end{aligned}
\end{equation}
where $L$ denotes the training horizon, \textit{i.e.}, the number of future steps used for multi-step supervision, while $\lambda_{\mathrm{lat}}$ and $\lambda_{\mathrm{pred}}$ balance latent consistency and physical space prediction. We empirically find that a moderate $L$ performs best, whereas overly short or long horizons degrade performance. As shown in Figure~\ref{fig:eigenvalue}, an appropriate training horizon keeps the eigenvalues of the learned Koopman operator $\mathcal{K}^{\Delta t}=\exp(\Delta tA)$ close to 1, with its spectral radius $\rho(\mathcal{K}^{\Delta t})$ only slightly greater than 1, which improves long-horizon stability during inference. 
% All components are trained end-to-end through the differentiable matrix exponential. 
At inference, $\mathcal{K}^{\tau}=\exp(\tau A)$ enables direct prediction at arbitrary horizons $\tau$ without recursive rollout or restriction to $\Delta t$.

\textbf{Training of Generative Correction.}
According to Theorem~\ref{them:ext_reverse}, finite-dimensional Koopman evolution induces a degradation distribution over latent states, from which the true latent distribution can be recovered through a corresponding reverse process. Building on this result, the objective is to train a conditional generative model to learn this reverse process using stochastic intermediate states constructed between reference latent states and Koopman predictions. Throughout Stage II, the encoder $E_\theta$, Koopman operator, and decoder $D_\theta$ remain frozen, while only the parameters of the correction model $\mathcal{C}_\phi$ are optimized.

With the model trained in Stage I, samples of the Koopman-induced degradation distribution $\widetilde{z}_K$ at target time $K$ can be obtained by $\{\widetilde{z}_K^{(i)}:\widehat{\mathcal{K}}^{K-i}z_i\}_{i<K}$, along with the corresponding ground-truth latent state $z_K$. Accordingly, the stochastic process to be learned is characterized by the following endpoint conditions, connecting the reference latent state to its corresponding degraded prediction:
% The generative correction model $\mathcal{C}\phi$ is then trained to learn the reverse process from the Koopman-induced degradation distribution to the true latent distribution
\begin{equation}
    Z_0=z_K=z_0,\quad Z_1=\widetilde z_K^{(i)}\sim q_{\mathrm{deg}}(\cdot|z_K).
\end{equation}
% For clarity, we omit the trajectory and temporal-pair superscripts and write
% \begin{equation}
% \widetilde{z}_K=\widetilde{z}_K^{\,i\rightarrow j},
% \quad z_0=z_j^\star,
% \quad\Delta t=\Delta t_{ij}.
% \end{equation}
% Here, $z_0$ denotes the clean target latent state rather than the initial state of the physical trajectory.
% We construct stochastic intermediate states using a conditional Brownian-bridge formulation. Let $\tau\in[0,1]$ denote the normalized sampling time, which is independent of the physical prediction interval $\Delta t=K-i$. The conditional distribution of an intermediate latent state is defined as
To learn the reverse process, we formulate a conditional Brownian-bridge between the two endpoint conditions $\widetilde{z}_K^{(i)}$ and $z_0$, providing a stochastic path for generative correction~\cite{li2023bbdm,wang2024score,xue2025bibbdm}. Let $s\in[0,1]$ denote the normalized sampling time. The conditional distribution of an intermediate latent state is then defined as:
\begin{equation}
q_{\mathrm{BB}}(z_s\mid z_0,\widetilde{z}_K)
=\mathcal{N}\left(z_s;(1-s)z_0+s\widetilde{z}_K,\delta_s I\right),
\end{equation}
where $\delta_s=2vs(1-s)$, and $v>0$ controls the variance of the stochastic perturbation. Specifically, intermediate states are sampled directly as:
\begin{equation}
z_s=(1-s)z_0+s\widetilde{z}_K
+\sqrt{\delta_s}\,\epsilon,\quad\epsilon\sim\mathcal{N}(0,I).
\end{equation}
This formulation anchors the process at the reference state when $s=0$ and at the degraded Koopman prediction when $s=1$, with stochastic perturbations introduced at intermediate times. It is notable that to ensure coverage of the inference starting point $s=1$, we sample $s$ from the discrete schedule $\{1/M,2/M,\cdots,1\}$ during training, where $M$ denotes the number of discretization steps. 

\begin{algorithm}[t]
  \caption{Pseudocode of KoopFlow Inference}
  \label{alg:koopflow_inference}
  \begin{algorithmic}
    \STATE {\bfseries Input:} $u_{n-K+1:n},\mu,\mathcal{B}$, horizon $H$, step size $h$, steps $M$
    \STATE $\widehat z^{(0)}=E_\theta(u_{n-K+1:n},\mu,\mathcal{B})$
    \STATE $N=\lfloor H/h\rfloor,\quad r=H\bmod h$
    \FOR{$j=1$ {\bfseries to} $N$}
        \STATE $\widetilde z^{(j)}=\exp(h\Delta t A)\widehat z^{(j-1)}$ (Stage~I)
        \STATE $\widehat z^{(j)}=\mathcal{C}_\phi^{(M)}(\widetilde z^{(j)},h\Delta t)$ (Stage~II)
    \ENDFOR
    \IF{$r>0$}
        \STATE $\widetilde z=\exp(r\Delta t A)\widehat z^{(N)}$
        \STATE $\widehat z=\mathcal{C}_\phi^{(M)}(\widetilde z,r\Delta t)$
    \ELSE
        \STATE $\widehat z=\widehat z^{(N)}$
    \ENDIF
    \STATE {\bfseries Output:} $\widehat u_{n+H}=D_\theta(\widehat z)$
  \end{algorithmic}
\end{algorithm}

\begin{table*}[t]
\centering
\caption{
Average NMSE on NS, RD, 2DHeat, and 2DEM. \textbf{Train} denotes forecast time steps within the training-time range 
% ; \textbf{All} denotes the full EM2D test horizon, 
and \textbf{OOD} denotes time steps beyond the
training-time range.
(\textbf{Bold}: best; \underline{Underline}: second best).
}
% \vspace{-3mm}
\label{tab:temporal_ood}
\small
\setlength{\tabcolsep}{4pt}
\resizebox{\textwidth}{!}{%
\begin{tabular}{lcccccccc}
\toprule
\multirow{2}{*}{Method}
& \multicolumn{2}{c}{Navier--Stokes}
& \multicolumn{2}{c}{Reaction--Diffusion}
& \multicolumn{2}{c}{2DHeat}
& \multicolumn{2}{c}{2DEM} \\
\cmidrule(lr){2-3}
\cmidrule(lr){4-5}
\cmidrule(lr){6-7}
\cmidrule(lr){8-9}
& Train & OOD
& Train & OOD
& Train & OOD
& Train & OOD \\
\midrule

DeepONet~\citep{lu2021deeponet}
& 2.632e-02 & 2.305e-01
& 5.396e-01 & 1.570e-01
& 2.089e-01 & 5.495e-02
& 1.219e+00 & 8.525e-01 \\

FNO~\citep{li2021fno}
& 8.096e-03 & \underline{1.154e-01}
& \textbf{6.612e-04} & 7.777e-03
& \textbf{3.297e-03} & \underline{1.292e-02}
& 1.136e-01 & 4.374e-01 \\

KNO~\citep{xiong2024kno}
& \underline{3.254e-03} & 1.940e+00
& 2.727e-03 & \underline{2.685e-03}
& 2.828e-02 & 1.451e-02
& 7.870e+00 & 1.255e+01 \\

LoLA~\citep{rozet2025lola}
& 4.648e-02 & 1.300e-01
& 2.043e-01 & 2.356e-02
& 2.590e-01 & 6.877e-02
& \underline{9.523e-02} & \underline{3.231e-01} \\

InfoKoop~\citep{cheng2026information}
& 2.976e-01 & 7.082e-01
& 5.499e-01 & 1.582e-01
& 6.170e-01 & 9.768e-02
& 1.145e+00 & 5.972e-01 \\

CFO~\citep{hou2026cfo}
& 5.313e-02 & 3.388e-01
& 2.460e-03 & 1.818e-01
& 3.923e-02 & 2.024e-01
& 2.252e+00 & 1.808e+00 \\

MERLIN~\citep{lu2026merlin}
& 2.097e-02 & 1.626e-01
& 6.533e-01 & 2.967e-01
& 3.811e-01 & 4.803e-02
& 5.002e-01 & 6.847e-01 \\

% % \midrule
% % IMP to $2^{nd}$ Best
% % & 83.94\% & 51.71\%
% % & 63.39\% & 9.721\%
% % & 6.341\% & 99.91\%
% % & 93.01\% & 93.37\% \\
\midrule
\textbf{KoopFlow (Ours)}
& \textbf{2.026e-03} & \textbf{5.572e-02}
& \underline{1.806e-03} & \textbf{2.424e-03}
& \underline{3.521e-03} & \textbf{1.174e-05}
& \textbf{9.107e-03} & \textbf{2.898e-02} \\

\bottomrule
\end{tabular}%
}
% \vspace{-5mm}
\vskip -0.1in
\end{table*}
Then the generative correction model takes the intermediate state, sampling time, Koopman prediction, and physical prediction interval as inputs and can be expressed as $\mathcal{C}_\phi\left(z_s,s,\widetilde{z}_K,T_i\right)$, where the conditioning variable $T_i=(K-i)\Delta t$ characterizes the physical forecasting horizon.
% The two time variables serve distinct purposes: $s$ specifies the position along the stochastic process, whereas $\Delta t$ characterizes the physical forecasting horizon. 
Following this, the model is trained to predict the total perturbation relative to the clean target:
\begin{equation}
\eta_s=z_s-z_0
=s(\widetilde{z}_K-z_0)
+\sqrt{\delta_s}\,\epsilon.
\end{equation}
Thus, the latent reconstruction objective is formulated as:
\begin{equation}
    \mathcal{L}_{\mathrm{den}}
    =\mathbb{E}\left[
    \left\|\mathcal{C}_\phi
    \left(z_s,s,\widetilde{z}_K,T_i\right)
    -\eta_s\right\|_2^2\right].
\end{equation}
This perturbation prediction objective explicitly trains the model to recover the target latent state from stochastic intermediate states conditioned on the degraded Koopman prediction.
Since $\eta_s=z_s-z_0$ and $\widehat{z}_0^\phi(s)=z_s-\mathcal{C}_\phi\left(z_s,s,\widetilde{z}_K,T_i\right)$, this objective is equivalent to minimizing the reconstruction error of the estimated clean endpoint:
\begin{equation}
\mathcal{L}_{\mathrm{den}}
=\mathbb{E}\left[
\left\|\widehat{z}_0^\phi(s)-z_0\right\|_2^2\right].
\end{equation}
% By subtracting the predicted perturbation, the estimated clean endpoint is consequently given by
% By learning to predict the total perturbation, the generative model is explicitly trained to recover the target latent state from stochastic intermediate states conditioned on the degraded Koopman prediction. Let $d_z$ denote the number of scalar elements in a latent state.
To further constrain reconstruction accuracy in the physical space, we decode the estimated state using the frozen decoder and define the physical-field reconstruction loss as:
\begin{equation}
\mathcal{L}_{\mathrm{field}}
=\mathbb{E}\left[
\left\|D_\theta\left(\widehat{z}_0^\phi(s)\right)-u_K
\right\|_2^2\right],
\end{equation}
where $u_K$ is the normalized ground-truth target field. Unlike latent supervision alone, this term directly evaluates the reconstructed physical field against the reference observation. Finally, the complete Stage II objective is:
\begin{equation}
\mathcal{L}_{\mathrm{Stage\,II}}
=\mathcal{L}_{\mathrm{den}}
+\lambda_{\mathrm{field}}\mathcal{L}_{\mathrm{field}}
\end{equation}
% \begin{equation}
% \begin{aligned}
% &\mathcal{L}_{\mathrm{Stage\,II}}
% =\mathcal{L}_{\mathrm{den}}
% +\lambda_{\mathrm{field}}\mathcal{L}_{\mathrm{field}}\\
% &=\mathbb{E}\Bigg[\left\|\mathcal{C}_\phi
% \left(z_s,s,\widetilde{z}_K,\Delta t\right)-(z_s-z_0)
% \right\|_2^2\\
% &+\lambda_{\mathrm{field}}
% \left\|D_\theta\left(z_s-\mathcal{C}_\phi
% \left(z_s,s,\widetilde{z}_K,\Delta t\right)
% \right)-v_j\right\|_2^2\Bigg].
% \end{aligned}
% \end{equation}
Here, $\lambda_{\mathrm{field}}\geq0$ balances latent reconstruction accuracy and physical-field fidelity. 
% Both terms use elementwise mean squared error without additional time-dependent weighting. Only $\phi$ is updated during optimization; the encoder, Koopman dynamics, and decoder remain frozen. Gradients are propagated through the decoder with respect to its latent input so that the physical-field loss can optimize the correction model.

\textbf{Inference of KoopFlow.} At inference time, we first obtain a degraded latent prediction using the Koopman model trained in Stage I. Since the reference target is unavailable, the correction process starts directly from this prediction:
\begin{equation}
z_{s_0}=\widetilde{z}_K,
\qquad s_0=1.
\end{equation}
For a decreasing sampling schedule $1=s_0>s_1>\cdots>s_M=0$,
the estimated clean endpoint at step $k$ is:
\begin{equation}
\label{eq:end_infer}
\widehat{z}_{0,k}
=z_{s_k}-\mathcal{C}_\phi
\left(z_{s_k},s_k,\widetilde{z}_K,T_i\right).
\end{equation}
We then update the intermediate state according to the Brownian-bridge transition:
\begin{equation}
\label{eq:bb_infer}
\begin{aligned}
a_k&=\frac{s_{k+1}}{s_k},\\
z_{s_{k+1}}
&=a_kz_{s_k}
+(1-a_k)\widehat{z}_{0,k}\\
&+
\sqrt{2v(s_k-s_{k+1})a_k}\,\xi_k,
\quad
\xi_k\sim\mathcal{N}(0,I).
\end{aligned}
\end{equation}
We denote the complete $M$-step reverse-bridge sampling
procedure defined by Eqs.~\ref{eq:end_infer}-\ref{eq:bb_infer} as
$\mathcal{C}_\phi^{(M)}(\widetilde{z},T)$, which maps
a degraded Koopman prediction $\widetilde{z}$ to its corrected
latent state over a prediction interval $T$.
% At each intermediate step, an independent Gaussian perturbation is introduced according to the transition above, while the final transition has zero conditional variance. 
Once the sampling process reaches the terminal step, the resulting latent state is decoded into the physical space to obtain the corrected prediction:
\begin{equation}
\widehat{u}_K=D_\theta(z_{s_M})=D_\theta(\widehat z_{K}).
\end{equation}
Notably, the inference schedule steps $M$ can differ from training and $M=50$ is chosen during inference. 
% For recursive trajectory prediction, the corrected latent state at the end of each prediction block serves as the anchor for subsequent Koopman propagation, with the physical prediction interval recomputed relative to the updated anchor. 
As shown in Algorithm~\ref{alg:koopflow_inference}, for long-term rollouts, this rectification procedure is applied periodically and the corrected latent state at the end of each prediction block serves as the anchor for subsequent Koopman propagation. Between correction checkpoints, intermediate states are propagated purely via the continuous Koopman dynamics. 
The two-stage framework decouples efficient prediction from trajectory recovery, with Koopman evolution enabling fast arbitrary-time forecasting and the stochastic bridge transporting degraded predictions toward the target latent trajectory distribution.
\section{Experiments}
\subsection{Experimental Setup}
\textbf{Datasets.} We evaluate KoopFlow on four benchmarks: Navier–Stokes (NS), reaction–diffusion (RD), 2DHeat, and 2D electromagnetic scattering (2DEM). Detailed dataset configurations are provided in Appendix~\ref{app:equations_datasets}.

\begin{figure*}[t]
    % \vspace{-3mm}
\begin{center}
    \includegraphics[width=0.85\textwidth]{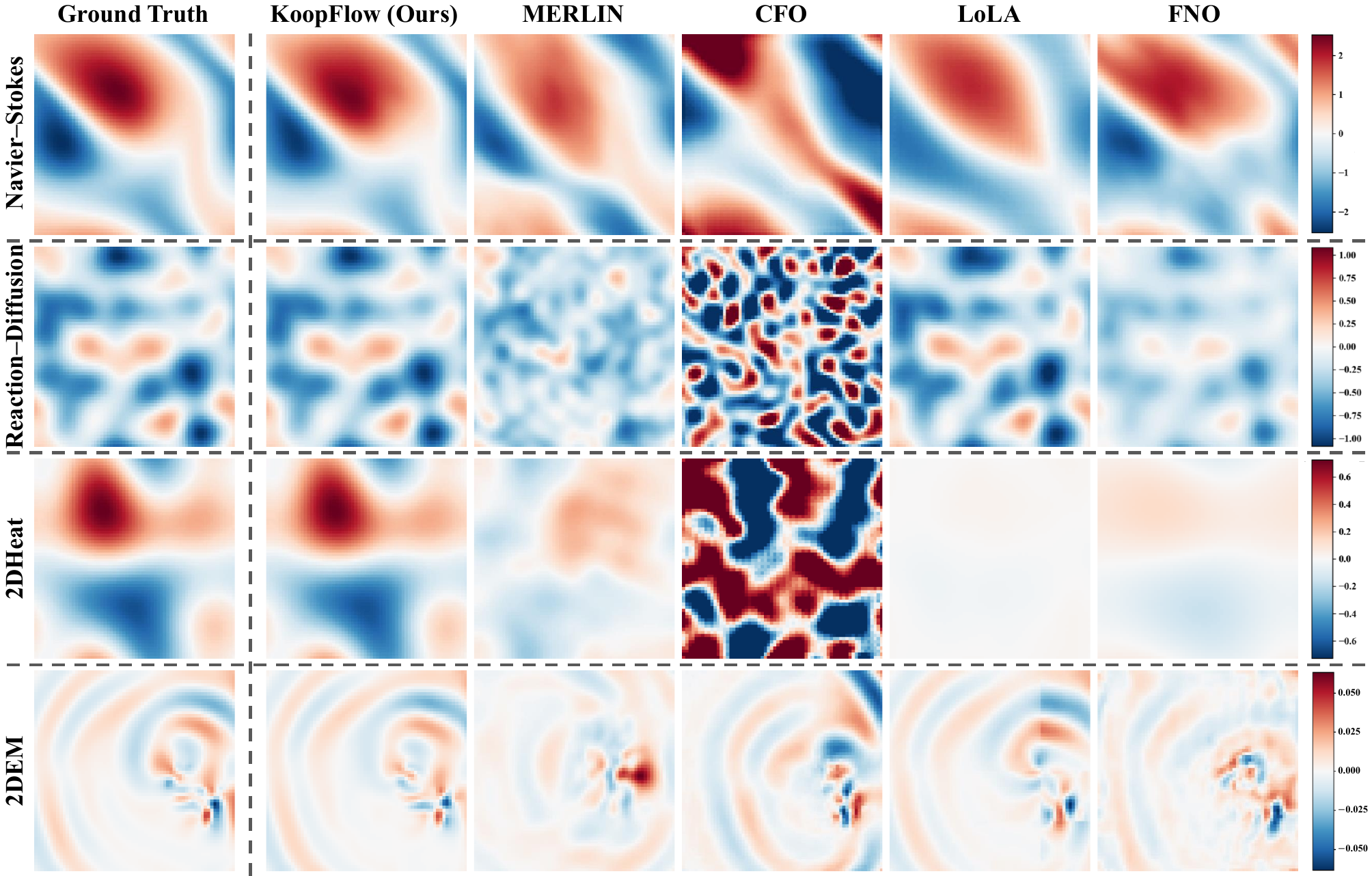}
    % \vspace{-5mm}
    \caption{Qualitative comparison of KoopFlow with other baselines on NS, RD, 2DHeat, and 2DEM at the final time. The dashed line separates reference results (left) from our method and four categories of representative methods (right).}
    % \vspace{-5mm}
    \label{fig:ns_traj451_final_time_field_error}
\end{center}
\end{figure*}
\textbf{Evaluation Metrics.} We assess forecasting performance using the normalized mean squared error (NMSE) and report the eigenvector condition number $\kappa(V)$ to quantify the conditioning of the learned Koopman eigenbasis~\citep{drmac2026koopmanschur}, where $V$ denotes the matrix of normalized eigenvectors of the latent Koopman generator. More details are provided in Appendix~\ref{app:metrics}.

\textbf{Baselines.}
We compare KoopFlow with several representative baseline methods spanning three complementary categories, including neural operators: DeepONet~\citep{lu2021deeponet} and FNO~\citep{li2021fno}; Koopman learning approaches: KNO~\citep{xiong2024kno}, InfoKoop~\citep{cheng2026information},
and MERLIN~\citep{lu2026merlin}; and generative models: LoLA~\citep{rozet2025lola}, CFO~\citep{hou2026cfo}. More training details of all these baseline models can be found in Appendix~\ref{app:benchmark_protocol}.
\begin{table*}[t]
\centering
\caption{
% Results on the 2DHeat generalization benchmark.
Average NMSE on 2DHeat Generalization Testset. \textbf{Seen} denotes parameter values encountered during training; \textbf{Inter} denotes unseen values within the training range; and \textbf{Extra (low/high)} denotes values below/above the training range respectively. (\textbf{Bold}: best; \underline{Underline}: second best).
}
\label{tab:heat2d_generalization}
\resizebox{0.80\textwidth}{!}{
\begin{tabular}{lccccc}
\toprule
Method
& Overall
& Seen
& Inter
& Extra (low)
& Extra (high) \\
\midrule
DeepONet~\citep{lu2021deeponet}
& 1.304e-01 & 1.050e-01 & 7.402e-02 & 3.544e-01 & 5.433e-02 \\

FNO~\citep{li2021fno}
& \underline{8.209e-03}
& \underline{3.147e-03}
& \underline{1.012e-03}
& \underline{3.924e-02}
& \underline{6.230e-04} \\

KNO~\citep{xiong2024kno}
& 2.126e-02 & 4.652e-03 & 4.575e-03 & 1.006e-01 & 8.441e-03 \\

LoLA~\citep{rozet2025lola}
& 1.619e-01 & 1.363e-01 & 8.142e-02 & 4.726e-01 & 3.622e-02 \\

InfoKoop~\citep{cheng2026information} % $^{\dagger}$
& 3.521e-01 & 3.223e-01 & 2.357e-01 & 8.010e-01 & 1.519e-01 \\

CFO~\citep{hou2026cfo}
& 1.225e-01 & 1.147e-01 & 9.620e-02 & 2.236e-01 & 8.011e-02 \\

MERLIN~\citep{lu2026merlin}
& 2.112e-01 & 1.781e-01 & 1.180e-01 & 5.673e-01 & 7.761e-02 \\
\midrule
\textbf{KoopFlow}
& \textbf{1.730e-03}
& \textbf{9.260e-04}
& \textbf{2.900e-04}
& \textbf{7.564e-03}
& \textbf{6.800e-05} \\
\bottomrule
\end{tabular}
}
\end{table*}
\begin{figure*}[t]
    \centering
    % \vspace{-3mm}
    \includegraphics[width=0.9\linewidth]{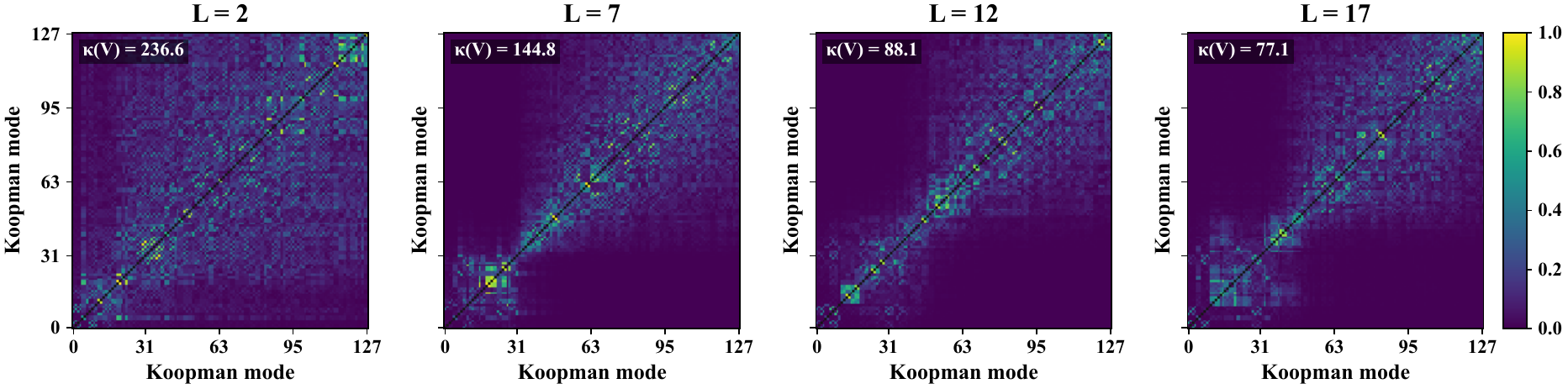}
    % \vspace{-3mm}
    \caption{Eigenvector overlap of Koopman representations under different training horizons on NS. The results from left to right correspond to progressively longer training window size.}
    % \vspace{-5mm}
    \label{fig:ns_eigenvector_overlap}
\end{figure*}

\subsection{Experimental Results}
% \subsection{Comparison with Baselines}

\textbf{Long-horizon forecasting performance.}
Table~\ref{tab:temporal_ood} compares KoopFlow against representative baselines on NS, RD, 2DHeat and 2DEM. Within the training-time range, KoopFlow remains highly competitive, achieving the best performance on NS and 2DEM, while attaining  the second-best performance on RD and 2DHeat. More importantly, KoopFlow consistently achieves the lowest NMSE across all four benchmarks in the temporal-OOD regime. Compared with the strongest baseline on each benchmark, KoopFlow reduces the OOD error by \(51.7\%\), \(9.7\%\), \(99.9\%\), and \(91.0\%\) on NS, RD, 2DHeat, and 2DEM. 
These results indicate that KoopFlow maintains strong predictive performance within the observed temporal range while exhibiting substantially improved temporal extrapolation beyond the training horizon.
Figure~\ref{fig:ns_traj451_final_time_field_error} also supports the observation that KoopFlow produces forecasts that better preserve the structures of dynamic systems, particularly at the final time. More visualization comparisons among KoopFlow and other baselines are shown in Appendix~\ref{app:more_qualitative_examples}

% Table~\ref{tab:heat2d_generalization} 
% % in Appendix~\ref{app:generalization} 
% further shows that KoopFlow achieves the lowest error across all evaluation regimes under unseen physical parameters. Compared to FNO, the strongest baseline across all five metrics, KoopFlow reduces the overall error by $78.9\%$, and improves the seen and interpolation errors by $70.6\%$ and $71.3\%$, respectively. The improvement is even more pronounced under parameter extrapolation, with error reductions of $80.7\%$ and $89.1\%$ for the low- and high-parameter regimes. These results demonstrate that KoopFlow exhibits consistent predictive accuracy across the same family of equations, even under previously unseen parameter settings, highlighting its strong generalization capability.
We further evaluate the generalization ability of KoopFlow to unseen physical parameters on 2DHeat. As shown in Table~\ref{tab:heat2d_generalization}, KoopFlow achieves the lowest error across all evaluation regimes, including seen parameters, interpolation within the training range, and extrapolation beyond the lower and upper boundaries of the training range. Compared with FNO, the strongest baseline across all five metrics, KoopFlow reduces the overall error by $78.9\%$, and improves the seen and interpolation errors by $70.6\%$ and $71.3\%$, respectively. The improvement is even more pronounced under parameter extrapolation, with error reductions of $80.7\%$ and $89.1\%$
for the low- and high-parameter regimes. These results demonstrate that KoopFlow exhibits consistent predictive accuracy across the same family of equations, even under previously unseen parameter settings, highlighting its strong generalization capability. Further analysis is provided in the Appendix~\ref{app:generalization}.

% \end{minipage}\hfill
% \begin{minipage}[t]{0.48\linewidth}

\begin{table}[t]
% \vspace{-3mm}
\centering
% \begin{minipage}[t]{0.48\linewidth}
% \vspace{0pt}
% \centering
\caption{Effect of recursive rollout steps. \(N\) denotes the number of recursive prediction segments over the given forecasting horizon.}
% \vspace{-3mm}
\label{tab:recursive_correction_ablation}
\small
\renewcommand{\arraystretch}{1.15}
\setlength{\tabcolsep}{2pt}
\begin{tabular*}{\linewidth}{@{\extracolsep{\fill}}lccc}
\toprule
Correction strategy & $N$ & NMSE $\downarrow$ & Reduction $\uparrow$ \\
\midrule
Direct & 1 & 5.408e{-1} & -- \\
Recursive & 3 & 1.814e{-1}   & 66.5\% \\
Recursive & 5 & 1.923e{-1} & 64.4\% \\
Recursive & 7 & 1.670e{-1}  & $\mathbf{69.1\%}$ \\
\bottomrule
\end{tabular*}
\vskip -0.1in
\end{table}

\textbf{Effect of recursive rollout steps.} 
To evaluate the effect of recursive rollout granularity, we vary the number of prediction segments $N$ while keeping the overall forecasting horizon fixed. 
% The results are reported in Table~\ref{tab:recursive_correction_ablation}, where $N=1$ corresponds to direct prediction without intermediate prediction.It is evident that recursive correction consistently outperforms direct correction over the same forecasting horizon, reducing NMSE by more than 64\%, with the best reduction of 69.1\% achieved at $N=7$.
As shown in Table~\ref{tab:recursive_correction_ablation}, $N=1$ denotes direct prediction with a single correction at the target time. Recursive correction consistently outperforms direct correction, reducing NMSE by over 64\% across all tested settings, with the best reduction of 69.1\% at $N=7$. This improvement arises because the continuous-time Koopman representation enables flexible temporal partitioning, allowing the recursive correction scheme to interleave Koopman evolution with generative correction. By shortening the uncorrected propagation intervals, this strategy mitigates the accumulation of approximation errors. Varying these intervals also exposes the generative model to different horizon-dependent degradation levels, consistent with the degradation distribution characterized in our theoretical framework. Notably, the improvement is not strictly monotonic with $N$, as $N=5$ yields a slightly higher NMSE than $N=3$, indicating that additional corrections do not necessarily improve accuracy. This non-monotonic behavior suggests a trade-off between mitigating propagation errors and introducing additional correction errors, highlighting the importance of selecting an appropriate correction frequency to balance predictive accuracy and computational cost.

\begin{table}[ht]
% \vspace{0pt}
\centering
\caption{Effect of Koopman training horizon. $\kappa(V)$ denotes the condition number of the Koopman eigenvector matrix, with smaller values indicating better numerical conditioning.}
% \vspace{-3mm}
\label{tab:ns_window_roll47}
\small
\renewcommand{\arraystretch}{1.15}
\setlength{\tabcolsep}{2pt}
\begin{tabular*}{\linewidth}{@{\extracolsep{\fill}}lccc}
\toprule
Horizon & $\kappa(V)$ & Koopman NMSE & KoopFlow NMSE \\
\midrule
$L=2$ & 236.6 & 4.690e{-1} & 5.523e{-2} \\
$L=7$ & 144.8 & 1.406e{-1} & 2.647e{-2} \\
$\mathbf{L=12}$ & 88.1 & \textbf{8.296e{-2}} & \textbf{2.489e{-2}} \\
$L=17$ & \textbf{77.1} & 8.564e{-2} & 2.976e{-2} \\
\bottomrule
\end{tabular*}
% \end{minipage}
% \vspace{-5mm}
\vskip -0.1in
\end{table}
\textbf{Effect of the Koopman training horizon.} 
Figure~\ref{fig:ns_eigenvector_overlap} and Table~\ref{tab:ns_window_roll47} both show that increasing the Koopman training horizon progressively improves the conditioning of the learned eigenbasis.
% , but does not lead to monotonically better forecasting accuracy.
In Figure~\ref{fig:ns_eigenvector_overlap}, it is evident that the eigenvector overlap matrices become increasingly concentrated along the diagonal as $L$ increases, with reduced off-diagonal overlap and clearer mode-wise correspondence. This trend is consistent with the improved conditioning of the learned eigenbasis, although results in Table~\ref{tab:ns_window_roll47} reveal that it does not necessarily translate into better forecasting accuracy.
The reported NMSE is averaged over the entire forecasting interval from time steps 4 to 50. As $L$ increases from $2$ to $7$ and $12$, $\kappa(V)$ decreases from $236.6$ to $144.8$ and $88.1$, while the prediction errors of both the Koopman backbone and KoopFlow
decrease substantially. The best forecasting performance is obtained at $L=12$. Increasing the horizon further to $L=17$ reduces $\kappa(V)$ to $77.1$, but slightly increases the prediction error. This observation suggests that better-conditioned Koopman representations alone do not necessarily translate into improved forecasting accuracy and that a moderate training horizon provides a better balance between latent dynamical conditioning and predictive fidelity.
% Importantly, KoopFlow consistently improves the prediction accuracy of the underlying Koopman backbone across all training horizons. Figure~\ref{fig:ns_eigenvector_overlap} further shows progressively reduced  overlap of eigenvector as $L$ increases.
Importantly, KoopFlow consistently improves the prediction accuracy of the underlying Koopman backbone across all training horizons. 
%% ------------------------------------------------
% \begin{table}[ht]
% \centering
% \captionof{table}{Ablation of rectification depth.}
% \label{tab:rectification_depth}
% \small
% % \renewcommand{\arraystretch}{1.5}
% % \setlength{\tabcolsep}{2pt}
% \begin{tabular}{c|cccc}
%     \toprule
%     $s_r$ & $L$=1 & $L$=2 & $L$=4 & $L$=8 \\
%     \midrule
%     0.6 & 2.350e-02	& 5.240e-02	& 7.929e-02	& 2.755e-01 \\
%     0.7 & 2.759e-02	& 5.784e-02	& 7.979e-02	& 2.746e-01 \\
%     0.8 & 5.509e-02	& 7.115e-02	& 8.486e-02	& 2.776e-01 \\
%     \bottomrule
% \end{tabular}
% \end{table}

% \textbf{Effect of rectification depth $s_r$.} Table~\ref{tab:rectification_depth} studies the effect of the Koopman propagation window $L$ and the rectification starting point $s_r$, where $s_r=0$ corresponds to pure noise and $s_r=1$ to clean data. As $L$ increases, comparable recovery increasingly favors noisier starting points (smaller $s_r$), while the overall error also grows substantially. This horizon-dependent shift empirically supports our theory that different Koopman propagation horizons induce distinct degradation levels, which collectively form a horizon-dependent degradation distribution, with longer horizons corresponding to more severe degradation.
%% ------------------------------------------------
% \begin{figure}[H]
    % \vspace{-3mm}
    % \centering
    % Left: Table
    % \begin{minipage}[t]{0.40\textwidth}
        % \vspace{0pt}
    % \end{minipage}
    % \hfill
    % Right: Figure
    % \begin{minipage}[t]{0.58\textwidth}
        % \vspace{0pt}
\begin{figure}[h]
\centering
\includegraphics[width=1\linewidth]{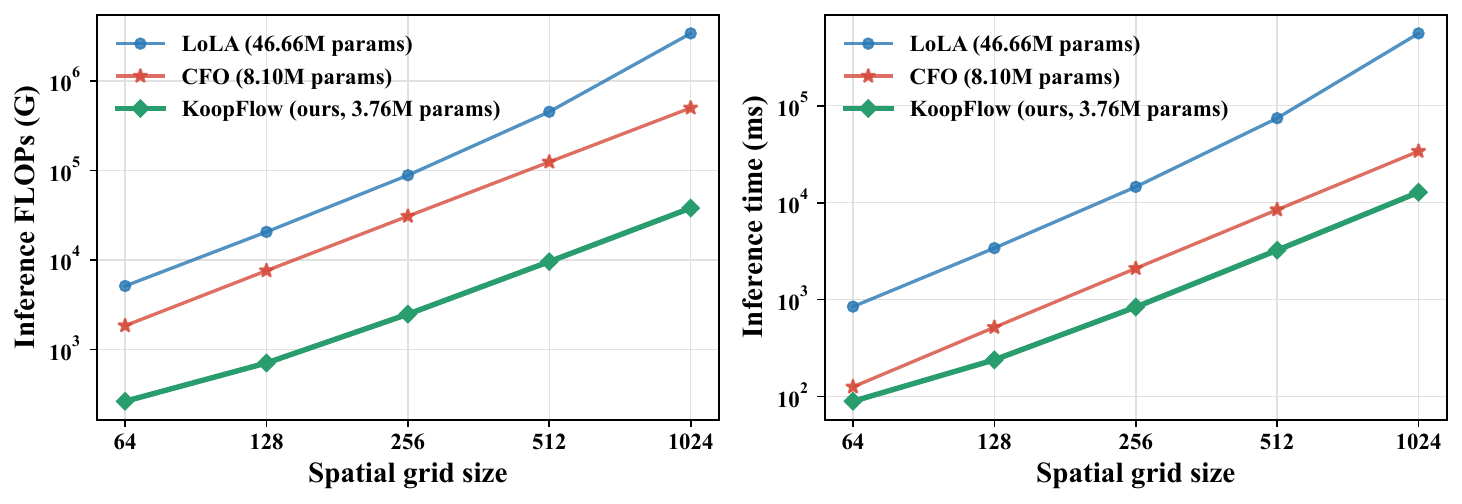}
% \vspace{-7mm}
\captionof{figure}{
     Comparison of inference FLOPs (left) and inference runtime (right) for KoopFlow, CFO, and LoLA across varying spatial grid resolutions.
}
\label{fig:lola_cfo_koopflow_inference_efficiency}
    % \end{minipage}
    % \vspace{-5mm}
\end{figure}

\textbf{Inference efficiency.}
% KoopFlow, CFO, and LoLA are iterative generative models, enabling a direct comparison of their computational costs. As shown in Figure~\ref{fig:lola_cfo_koopflow_inference_efficiency}, KoopFlow has $2.2\times$ fewer parameters than CFO and $12.4\times$ fewer than LoLA. At the $64\times64$ spatial grid size, it requires $7.0\times$ and $19.4\times$ less computation and is $1.4\times$ and $9.5\times$ faster than CFO and LoLA. Moreover, these efficiency gains become increasingly pronounced as the spatial grid size grows, demonstrating that KoopFlow is a more efficient and more stable generative model.
We compare the inference efficiency of KoopFlow with CFO and LoLA, as all three methods employ iterative generative inference. As shown in Figure~\ref{fig:lola_cfo_koopflow_inference_efficiency}, KoopFlow has $2.2\times$ fewer parameters than CFO and $12.4\times$ fewer than LoLA. At the $64\times64$ spatial grid resolution, it requires $7.0\times$ and $19.4\times$ fewer FLOPs and achieves $1.4\times$ and $9.5\times$ faster inference than CFO and LoLA respectively. Moreover, these computational and runtime advantages become increasingly pronounced as the spatial grid size grows, demonstrating the favorable scalability of KoopFlow to higher-resolution problems. Together with the improved predictive stability demonstrated in our previous experiments, these results highlight KoopFlow as an efficient and stable generative framework.

\section{Conclusion}
\label{conclusion}
We propose KoopFlow, a probabilistic Koopman learning framework for modeling the accumulated effects of finite-dimensional approximation and closure errors. 
Our key theoretical insight is to reinterpret finite-dimensional Koopman error accumulation as a horizon-dependent degradation. Based on this view, KoopFlow couples efficient linear Koopman evolution with generative correction to recover degraded latent dynamics, effectively mitigating error accumulation during long-term rollouts. Experiments across diverse dynamical systems demonstrate improved extrapolation accuracy and stability while maintaining a favorable accuracy-efficiency trade-off.

%%%%%%%%%%%%%%%%%%%%%%%%%%%%%%%%%%%%%%%%%%%%%%%%%%%%%%%%%%%%%%%%%%%%%%%%%%%%%%%
% In the unusual situation where you want a paper to appear in the
% references without citing it in the main text, use \nocite
\nocite{langley00}

\bibliography{example_paper}

@article{lagemann2026hydrogym,
  title={The HydroGym reinforcement learning platform for fluid dynamics},
  author={Lagemann, Christian and Mokbel, Sajeda and Gondrum, Miro and R{\"u}ttgers, Mario and Wang, Yuning and Su{\'a}rez, Pol and Paehler, Ludger and Bezgin, Deniz A and Buhendwa, Aaron B and Callaham, Jared L and others},
  journal={Nature},
  pages={1--8},
  year={2026},
  publisher={Nature Publishing Group UK London}
}

@article{bi2023accurate,
  title={Accurate medium-range global weather forecasting with 3D neural networks},
  author={Bi, Kaifeng and Xie, Lingxi and Zhang, Hengheng and Chen, Xin and Gu, Xiaotao and Tian, Qi},
  journal={Nature},
  volume={619},
  number={7970},
  pages={533--538},
  year={2023},
  publisher={Nature Publishing Group UK London}
}

@inproceedings{cao2026physics,
  title={Physics-Augmented Graph Transformer for Multi-step Time-Domain Maxwell's Equations in Electromagnetic Scattering},
  author={Cao, Huilin and Sun, Haoran and Liang, Jingwei and Li, Maokun and Ren, Yuxiang},
  booktitle={Proceedings of the 32nd ACM SIGKDD Conference on Knowledge Discovery and Data Mining V. 2},
  pages={10532--10542},
  year={2026}
}

@article{koopman1931hamiltonian,
  title={Hamiltonian systems and transformation in Hilbert space},
  author={Koopman, Bernard O},
  journal={Proceedings of the National Academy of Sciences},
  volume={17},
  number={5},
  pages={315--318},
  year={1931}
}

@article{chorin2000optimal,
  title   = {Optimal Prediction and the Mori--Zwanzig Representation of
             Irreversible Processes},
  author  = {Chorin, Alexandre J. and Hald, Ole H. and Kupferman, Raz},
  journal = {Proceedings of the National Academy of Sciences},
  volume  = {97},
  number  = {7},
  pages   = {2968--2973},
  year    = {2000},
  doi     = {10.1073/pnas.97.7.2968}
}

@article{mezic2005spectral,
  title={Spectral properties of dynamical systems, model reduction and decompositions},
  author={Mezi{\'c}, Igor},
  journal={Nonlinear Dynamics},
  volume={41},
  number={1},
  pages={309--325},
  year={2005},
  publisher={Springer}
}

@article{brunton2016koopman,
  title   = {Koopman Invariant Subspaces and Finite Linear Representations
             of Nonlinear Dynamical Systems for Control},
  author  = {Brunton, Steven L. and Brunton, Bingni W. and
             Proctor, Joshua L. and Kutz, J. Nathan},
  journal = {PLOS ONE},
  volume  = {11},
  number  = {2},
  pages   = {e0150171},
  year    = {2016},
  doi     = {10.1371/journal.pone.0150171}
}

@article{lusch2018deep,
  title   = {Deep Learning for Universal Linear Embeddings of Nonlinear Dynamics},
  author  = {Lusch, Bethany and Kutz, J. Nathan and Brunton, Steven L.},
  journal = {Nature Communications},
  volume  = {9},
  pages   = {4950},
  year    = {2018},
  doi     = {10.1038/s41467-018-07210-0}
}

@inproceedings{azencot2020forecasting,
  title={Forecasting sequential data using consistent Koopman autoencoders},
  author={Azencot, Omri and Erichson, N Benjamin and Lin, Vanessa and Mahoney, Michael},
  booktitle={International conference on machine learning},
  pages={475--485},
  year={2020},
  organization={PMLR}
}

@article{lin2021mori,
  title   = {Data-Driven Learning for the Mori--Zwanzig Formalism:
             A Generalization of the Koopman Learning Framework},
  author  = {Lin, Yen Ting and Tian, Yifeng and Livescu, Daniel and Anghel, Marian},
  journal = {SIAM Journal on Applied Dynamical Systems},
  volume  = {20},
  number  = {4},
  pages   = {2558--2601},
  year    = {2021},
  doi     = {10.1137/21M1401759}
}

@inproceedings{li2021fno,
  title     = {Fourier Neural Operator for Parametric Partial Differential Equations},
  author    = {Li, Zongyi and Kovachki, Nikola and Azizzadenesheli, Kamyar
               and Liu, Burigede and Bhattacharya, Kaushik and
               Stuart, Andrew and Anandkumar, Anima},
  booktitle = {International Conference on Learning Representations},
  year      = {2021},
  url       = {https://openreview.net/forum?id=c8P9NQVtmnO}
}

@article{lu2021deeponet,
  title   = {Learning Nonlinear Operators via {DeepONet} Based on the
             Universal Approximation Theorem of Operators},
  author  = {Lu, Lu and Jin, Pengzhan and Pang, Guofei and
             Zhang, Zhongqiang and Karniadakis, George Em},
  journal = {Nature Machine Intelligence},
  volume  = {3},
  pages   = {218--229},
  year    = {2021},
  doi     = {10.1038/s42256-021-00302-5}
}

@article{brunton2021modern,
  title = {Modern Koopman Theory for Dynamical Systems},
  author = {Brunton, Steven L. and Budi{\v s}i{\'c}, Marko and Kaiser, Eurika and Kutz, J. Nathan},
  journal = {SIAM Review},
  volume = {64},
  number = {2},
  pages = {229-340},
  year = {2022},
  publisher={SIAM}
}

@book{mauroy2020koopman,
  title={Koopman Operator in Systems and Control: Concepts, Methodologies, and Applications},
  author={Mauroy and Mezi{\'c}, Igor and Susuki, Yoshihiko},
  year={2020},
  publisher={Springer International Publishing}
}

@article{xiong2024kno,
  title   = {Koopman Neural Operator as a Mesh-Free Solver of Non-Linear
             Partial Differential Equations},
  author  = {Xiong, Wei and Huang, Xiaomeng and Zhang, Ziyang and
             Deng, Ruixuan and Sun, Pei and Tian, Yang},
  journal = {Journal of Computational Physics},
  volume  = {513},
  pages   = {113194},
  year    = {2024},
  doi     = {10.1016/j.jcp.2024.113194}
}

@inproceedings{lu2026merlin,
  title     = {Interpretable Functional Koopman Learning with Non-Markovian Closure for Spatiotemporal Systems},
  author    = {Lu, Wanfeng and Ma, He and Lin, Wei and Zhu, Qunxi},
  booktitle = {Proceedings of the 43rd International Conference on Machine Learning},
  year      = {2026},
  url       = {https://openreview.net/forum?id=lquDiBCgNJ}
}

@inproceedings{rozet2025lola,
  title     = {Lost in Latent Space: An Empirical Study of Latent Diffusion Models for Physics Emulation},
  author    = {Rozet, Fran{\c{c}}ois and Ohana, Ruben and McCabe, Michael and Louppe, Gilles and Lanusse, Fran{\c{c}}ois and Ho, Shirley},
  booktitle = {Advances in Neural Information Processing Systems},
  year      = {2025},
  url       = {https://openreview.net/forum?id=xoNrbfbekM}
}

@inproceedings{hou2026cfo,
  title     = {{CFO}: Learning Continuous-Time {PDE} Dynamics via Flow-Matched Neural Operators},
  author    = {Hou, Xianglong and Huang, Xinquan and Perdikaris, Paris},
  booktitle = {International Conference on Learning Representations},
  year      = {2026},
  url       = {https://openreview.net/forum?id=IQhaeSzyup}
}

@inproceedings{cheng2026information,
  title     = {Information Shapes Koopman Representation},
  author    = {Cheng, Xiaoyuan and Yuan, Wenxuan and Yang, Yiming and Zhang, Yuanzhao and Cheng, Sibo and He, Yi and Sun, Zhuo},
  booktitle = {International Conference on Learning Representations},
  year      = {2026},
  url       = {https://openreview.net/forum?id=Szh0ELyQxL}
}

@article{gupta2025mzae,
  title   = {Mori--Zwanzig Latent Space Koopman Closure for Nonlinear Autoencoder},
  author  = {Gupta, Priyam and Schmid, Peter J. and Sipp, Denis
             and Sayadi, Taraneh and Rigas, Georgios},
  journal = {Proceedings of the Royal Society A},
  volume  = {481},
  number  = {2313},
  pages   = {20240259},
  year    = {2025},
  doi     = {10.1098/rspa.2024.0259}
}

@article{drmac2026koopmanschur,
  title={A Data Driven Koopman--Schur Decomposition for Computational Analysis of Nonlinear Dynamics},
  author={Drma{\v{c}}, Zlatko and Mezi{\'c}, Igor},
  journal={SIAM Journal on Scientific Computing},
  volume={48},
  number={3},
  pages={A1430--A1461},
  year={2026},
  doi={10.1137/24M1672584}
}

@article{koehler2024apebench,
  title={{APEBench}: A benchmark for autoregressive neural emulators of pdes},
  author={Koehler, Felix and Niedermayr, Simon and Westermann, R{\"u}diger and Thuerey, Nils},
  journal={Advances in Neural Information Processing Systems},
  volume={37},
  pages={120252--120310},
  year={2024}
}

@article{takeishi2017learning,
  title={Learning Koopman invariant subspaces for dynamic mode decomposition},
  author={Takeishi, Naoya and Kawahara, Yoshinobu and Yairi, Takehisa},
  journal={Advances in neural information processing systems},
  volume={30},
  year={2017}
}

@article{mori1965transport,
  title={Transport, collective motion, and Brownian motion},
  author={Mori, Hazime},
  journal={Progress of theoretical physics},
  volume={33},
  number={3},
  pages={423--455},
  year={1965},
  publisher={Oxford University Press}
}

@article{zwanzig1973nonlinear,
  title={Nonlinear generalized Langevin equations},
  author={Zwanzig, Robert},
  journal={Journal of Statistical Physics},
  volume={9},
  number={3},
  pages={215--220},
  year={1973},
  publisher={Springer}
}

@article{haseli2021learning,
  title={Learning Koopman eigenfunctions and invariant subspaces from data: Symmetric subspace decomposition},
  author={Haseli, Masih and Cort{\'e}s, Jorge},
  journal={IEEE Transactions on Automatic Control},
  volume={67},
  number={7},
  pages={3442--3457},
  year={2021},
  publisher={IEEE}
}

@article{huang2024diffusionpde,
  title={Diffusionpde: Generative pde-solving under partial observation},
  author={Huang, Jiahe and Yang, Guandao and Wang, Zichen and Park, Jeong Joon},
  journal={Advances in Neural Information Processing Systems},
  volume={37},
  pages={130291--130323},
  year={2024}
}

@inproceedings{liu2025diff,
  title={Diff-PIC: Revolutionizing particle-in-cell nuclear fusion simulation with diffusion models},
  author={Liu, Chuan and Wu, Chunshu and Chen, Mingkai and Liang, James and Li, Ang and Huang, Michael and Ren, Chuang and Wu, Yingnian and Liu, Dongfang and Geng, Tong and others},
  booktitle={International Conference on Learning Representations},
  volume={2025},
  pages={71526--71539},
  year={2025}
}

@article{kohl2026benchmarking,
  title={Benchmarking autoregressive conditional diffusion models for turbulent flow simulation},
  author={Kohl, Georg and Chen, Li-Wei and Thuerey, Nils},
  journal={Neural Networks},
  pages={108641},
  year={2026},
  publisher={Elsevier}
}

@inproceedings{holzschuh2026p3d,
  title={P3d: Highly scalable 3d neural surrogates for physics simulations with global context},
  author={Holzschuh, Benjamin and Kohl, Georg and Redinger, Florian and Thuerey, Nils},
  booktitle={The Fourteenth International Conference on Learning Representations},
  year={2026}
}

@article{du2024conditional,
  title={Conditional neural field latent diffusion model for generating spatiotemporal turbulence},
  author={Du, Pan and Parikh, Meet Hemant and Fan, Xiantao and Liu, Xin-Yang and Wang, Jian-Xun},
  journal={Nature Communications},
  volume={15},
  number={1},
  pages={10416},
  year={2024},
  publisher={Nature Publishing Group UK London}
}

@inproceedings{zhou2025text2pde,
  title={Text2pde: Latent diffusion models for accessible physics simulation},
  author={Zhou, Anthony and Li, Zijie and Schneier, Michael and Buchanan, John and Barati Farimani, Amir},
  booktitle={International Conference on Learning Representations},
  volume={2025},
  pages={46344--46373},
  year={2025}
}

@article{nayak2025temporally,
  title={Temporally consistent Koopman autoencoders for forecasting dynamical systems},
  author={Nayak, Indranil and Chakrabarti, Ananda and Kumar, Mrinal and Teixeira, Fernando L and Goswami, Debdipta},
  journal={Scientific Reports},
  volume={15},
  number={1},
  pages={22127},
  year={2025},
  publisher={Nature Publishing Group UK, London}
}

@inproceedings{
shikhman2026semigroup,
title={Semigroup Consistency as a Diagnostic for Learned Physics Simulators},
author={Lennon J. Shikhman},
booktitle={ICML 2026 Workshop on AI for Physics},
year={2026},
url={https://openreview.net/forum?id=MeAFOZnrvM}
}

@article{shysheya2024conditional,
  title={On conditional diffusion models for PDE simulations},
  author={Shysheya, Aliaksandra and Diaconu, Cristiana and Bergamin, Federico and Perdikaris, Paris and Hern{\'a}ndez-Lobato, Jos{\'e} M and Turner, Richard E and Mathieu, Emile},
  journal={Advances in Neural Information Processing Systems},
  volume={37},
  pages={23246--23300},
  year={2024}
}

@article{bu2026sparsely,
  title={Sparsely-Supervised Data Assimilation via Physics-Informed Schr$\backslash$" odinger Bridge},
  author={Bu, Dohyun and Kim, Chanho and Choi, Seokun and Lee, Jong-Seok},
  journal={arXiv preprint arXiv:2603.22319},
  year={2026}
}

@inproceedings{li2023bbdm,
  title={Bbdm: Image-to-image translation with brownian bridge diffusion models},
  author={Li, Bo and Xue, Kaitao and Liu, Bin and Lai, Yu-Kun},
  booktitle={2023 IEEE/CVF Conference on Computer Vision and Pattern Recognition (CVPR)},
  pages={1952--1961},
  year={2023},
  organization={IEEE}
}

@article{liu20232,
  title={I $^{2}$ SB: Image-to-Image Schr$\backslash$" odinger Bridge},
  author={Liu, Guan-Horng and Vahdat, Arash and Huang, De-An and Theodorou, Evangelos A and Nie, Weili and Anandkumar, Anima},
  journal={arXiv preprint arXiv:2302.05872},
  year={2023}
}

@inproceedings{wang2024score,
  title={Score-based image-to-image brownian bridge},
  author={Wang, Peiyong and Xiao, Bohan and He, Qisheng and Glide-Hurst, Carri and Dong, Ming},
  booktitle={Proceedings of the 32nd ACM international conference on multimedia},
  pages={10765--10773},
  year={2024}
}

@inproceedings{lan2025schrodinger,
  title={When Schr{\"o}dinger bridge meets real-world image dehazing with unpaired training},
  author={Lan, Yunwei and Cui, Zhigao and Luo, Xin and Liu, Chang and Wang, Nian and Zhang, Menglin and Su, Yanzhao and Liu, Dong},
  booktitle={2025 IEEE/CVF International Conference on Computer Vision (ICCV)},
  pages={8756--8765},
  year={2025},
  organization={IEEE}
}

@article{de2021diffusion,
  title={Diffusion schr{\"o}dinger bridge with applications to score-based generative modeling},
  author={De Bortoli, Valentin and Thornton, James and Heng, Jeremy and Doucet, Arnaud},
  journal={Advances in neural information processing systems},
  volume={34},
  pages={17695--17709},
  year={2021}
}

@inproceedings{wang2021deep,
  title={Deep generative learning via schr{\"o}dinger bridge},
  author={Wang, Gefei and Jiao, Yuling and Xu, Qian and Wang, Yang and Yang, Can},
  booktitle={International conference on machine learning},
  pages={10794--10804},
  year={2021},
  organization={PMLR}
}

@article{xue2025bibbdm,
  title={Bibbdm: Bidirectional image translation with brownian bridge diffusion models},
  author={Xue, Kaitao and Li, Bo and Liu, Ziyi and He, Zhifen and Liu, Bin and Zhang, Congxuan and Lai, Yu-Kun},
  journal={IEEE Transactions on Pattern Analysis and Machine Intelligence},
  year={2025},
  publisher={IEEE}
}

@article{yue2023image,
  title={Image restoration through generalized ornstein-uhlenbeck bridge},
  author={Yue, Conghan and Peng, Zhengwei and Ma, Junlong and Du, Shiyan and Wei, Pengxu and Zhang, Dongyu},
  journal={arXiv preprint arXiv:2312.10299},
  year={2023}
}
\bibliographystyle{icml2026}

%%%%%%%%%%%%%%%%%%%%%%%%%%%%%%%%%%%%%%%%%%%%%%%%%%%%%%%%%%%%%%%%%%%%%%%%%%%%%%%
%%%%%%%%%%%%%%%%%%%%%%%%%%%%%%%%%%%%%%%%%%%%%%%%%%%%%%%%%%%%%%%%%%%%%%%%%%%%%%%
% APPENDIX
%%%%%%%%%%%%%%%%%%%%%%%%%%%%%%%%%%%%%%%%%%%%%%%%%%%%%%%%%%%%%%%%%%%%%%%%%%%%%%%
%%%%%%%%%%%%%%%%%%%%%%%%%%%%%%%%%%%%%%%%%%%%%%%%%%%%%%%%%%%%%%%%%%%%%%%%%%%%%%%
\newpage
\appendix
\onecolumn
\section{Theoretical Proofs}
\subsection{Proof of Lemma~\ref{lem:err_accumulation}}
\label{app:proof_lamma1}

\textbf{Lemma~\ref{lem:err_accumulation} (Accumulation of Koopman Approximation Error).}
\textit{
For any historical state $\psi_i\in\mathcal H$, define the $n$-step rollout deviation under the perturbed Koopman operator as $e_n^{(i)}=\widehat{\mathcal K}_{\mathcal H}^{n}\psi_i-\mathcal K^{n}\psi_i$. As $\|\Delta\mathcal K\|\to 0$, the accumulated error can be expressed as: 
\[
e_n^{(i)}=\sum_{j=0}^{n-1}\mathcal K^{n-1-j}
\Delta\mathcal K\mathcal K^j \psi_i+\mathcal O(\|\Delta\mathcal K\|^2\|\psi_i\|),
\]
}
% \[
%     e_n^{(i)}=\widehat{\mathcal K}_{\mathcal H}^{n}\psi_i-\mathcal K^{n}\psi_i .
% \]
% \textit{
% When $\|\Delta\mathcal K\|$ is sufficiently small, the accumulated error can be approximated as
% }
% \[
% e_n^{(i)}=\sum_{j=0}^{n-1}\mathcal K^{n-1-j}
% \Delta\mathcal K\mathcal K^j \psi_i+\mathcal O(\|\Delta\mathcal K\|^2).
% \]
\textbf{Proof of Lemma~\ref{lem:err_accumulation}.}
Let
\[
\widehat{\mathcal K}_{\mathcal H}=\mathcal K+\Delta\mathcal K.
\]
By definition
\[
E_n=(\mathcal K+\Delta\mathcal K)^n-\mathcal K^n.
\]
then it can be expanded as follows:
\[
\begin{aligned}
E_n
&=(\mathcal K+\Delta\mathcal K)^{n-1}(\mathcal K+\Delta\mathcal K)-\mathcal K^n\\
&=E_{n-1}\mathcal K+(\mathcal K+\Delta\mathcal K)^{n-1}\Delta\mathcal K.
\end{aligned}
\]
Applying this recursion iteratively and retaining the first-order
terms of $\Delta\mathcal K$ gives
\[
E_n=\sum_{j=0}^{n-1}\mathcal K^{n-1-j}\Delta\mathcal K\mathcal K^j
+O(\|\Delta\mathcal K\|^2).
\]
Multiplying both sides by $\psi_i$ yields
\[
e_n^{(i)}=E_n \psi_i=\sum_{j=0}^{n-1}\mathcal K^{n-1-j}\Delta\mathcal K\mathcal K^j \psi_i
+\mathcal O(\|\Delta\mathcal K\|^2\|\psi_i\|),
\]
which proves the lemma.
\hfill $\square$

This lemma establishes that finite-dimensional approximations of the Koopman operator inevitably introduce errors that accumulate during temporal evolution, leading to increasing deviations from the exact Koopman dynamics over nonzero prediction horizons.
% ------------------------------------
\subsection{Proof of Proposition~\ref{prop:koop_roll_err}}
\label{app:proof_prop1}
\textbf{Proposition~\ref{prop:koop_roll_err} (Horizon-dependent Koopman Rollout Error).}
\textit{
% Assume that the Koopman operator is bounded as $\|\mathcal{K}\|\leq\rho$, and the finite-dimensional approximation error satisfies $\|\Delta\mathcal{K}\|\leq\epsilon$. The $n$-step rollout deviation introduced by the finite-dimensional
% Koopman approximation satisfies:
% \[
%     \|e_n^{(i)}\|\leq n\rho^{n-1}\epsilon\|\psi_i\|+\mathcal{O}(\epsilon^2).
% \]
Assume that the exact Koopman evolution operator is bounded as $\|\mathcal{K}\|\leq\rho$, and that the lifted finite-dimensional approximation satisfies $\|\Delta\mathcal{K}\|\leq\epsilon$.
For any reference state $\psi_i$ and integer $n\geq1$, the rollout error $e_n^{(i)}= \widehat{\mathcal K}_{\mathcal H}^{\,n} \psi_i-\mathcal K^n\psi_i$ satisfies:
\[
\|e_n^{(i)}\|\leq\left[(\rho+\epsilon)^n-\rho^n\right]\|\psi_i\|.
\]
In particular, for fixed $n$ and $\epsilon\to0$,
\[
\|e_n^{(i)}\|\leq\left[n\rho^{n-1}\epsilon+\mathcal O_n(\epsilon^2)\right]\|\psi_i\|.
\]
}
% \[
%     \|e_n^{(i)}\|\leq n\rho^{n-1}\epsilon\|\psi_i\|+\mathcal{O}(\epsilon^2).
% \]
\textbf{Proof of Proposition~\ref{prop:koop_roll_err}.}
% According to Lemma~\ref{lem:err_accumulation}, the $n$-step rollout deviation can be expressed as
% \[
% e_n^{(i)}=\sum_{j=0}^{n-1}\mathcal K^{n-1-j}\Delta\mathcal K\mathcal K^j \psi_i
% +\mathcal O(\|\Delta\mathcal K\|^2).
% \]
% Taking the norm on both sides and applying the sub-multiplicative property of operator norms, we obtain
% \[
% \begin{aligned}
% \|e_n^{(i)}\|
% &\leq\sum_{j=0}^{n-1}\left\|\mathcal K^{n-1-j}\Delta\mathcal K\mathcal K^j \psi_i\right\|
% +\mathcal O(\|\Delta\mathcal K\|^2)
% \\
% &\leq\sum_{j=0}^{n-1}\|\mathcal K\|^{n-1-j}\|\Delta\mathcal K\|\|\mathcal K\|^j\|\psi_i\|
% +\mathcal O(\epsilon^2).
% \end{aligned}
% \]
% By the boundedness assumption \(\|\mathcal K\|\leq\rho\) and \(\|\Delta\mathcal K\|\leq\epsilon\), we have
% \[
% \begin{aligned}
% \|e_n^{(i)}\|
% &\leq\sum_{j=0}^{n-1}\rho^{n-1-j}\epsilon\rho^j\|\psi_i\|
% +\mathcal O(\epsilon^2)
% \\
% &=n\rho^{n-1}\epsilon\|\psi_i\|
% +\mathcal O(\epsilon^2),
% \end{aligned}
% \]
% which completes the proof.
Recall that
\[
\widehat{\mathcal K}_{\mathcal H}
=\mathcal K+\Delta\mathcal K.
\]
Using the telescoping identity, the difference between the
$n$-step evolutions can be decomposed as
\[
\widehat{\mathcal K}_{\mathcal H}^{\,n}-\mathcal K^n
=
\sum_{j=0}^{n-1}
\widehat{\mathcal K}_{\mathcal H}^{\,n-1-j}
\Delta\mathcal K\mathcal K^j.
\]
Applying the submultiplicativity of the operator norm and the
assumptions $\|\mathcal K\|\leq\rho$ and
$\|\Delta\mathcal K\|\leq\epsilon$, we obtain
\[
\begin{aligned}
\|e_n^{(i)}\|
&\leq
\sum_{j=0}^{n-1}
\|\widehat{\mathcal K}_{\mathcal H}\|^{n-1-j}
\|\Delta\mathcal K\|
\|\mathcal K\|^j
\|\psi_i\|\\
&\leq
\epsilon\|\psi_i\|
\sum_{j=0}^{n-1}
(\rho+\epsilon)^{n-1-j}\rho^j\\
&=
\left[(\rho+\epsilon)^n-\rho^n\right]
\|\psi_i\|.
\end{aligned}
\]
Here, we use
$\|\widehat{\mathcal K}_{\mathcal H}\|
\leq\|\mathcal K\|+\|\Delta\mathcal K\|
\leq\rho+\epsilon$.
This establishes the first inequality.

Furthermore, for fixed $n$ and $\epsilon\to0$, the binomial
expansion gives
\[
(\rho+\epsilon)^n-\rho^n
=n\rho^{n-1}\epsilon+\mathcal O_n(\epsilon^2).
\]
Consequently,
\[
\|e_n^{(i)}\|\leq\left[n\rho^{n-1}\epsilon
+\mathcal O_n(\epsilon^2)\right]\|\psi_i\|.
\]
This completes the proof.
\hfill $\square$

This proposition establishes a horizon-dependent bound on Koopman approximation errors, showing how these errors may be amplified during repeated evolution depending on the operator's growth properties.
\subsection{Proof of Theorem~\ref{them:ext_reverse}}
\label{app:proof_theo1}

\textbf{Theorem~\ref{them:ext_reverse} (Existence of Koopman-induced Reverse Process).}
\textit{
Given the Koopman-induced degradation distribution $q_{\mathrm{deg}}(\widetilde z_K|z_{0:K})$ in Definition~\ref{def:koop_dist}, there exists a continuous stochastic process $\{Z_s\}_{s\in[0,1]}$ satisfying:
\[
    Z_0=z_K,\qquad Z_1=\widetilde z_K,
    \quad\widetilde z_K\sim q_{\mathrm{deg}}(\cdot|z_{0:K}).
\]
Moreover, its time-reversed process $\bar Z_s:=Z_{1-s}$ defines a continuous reverse process from the degraded state to the target latent state, satisfying $\bar Z_0=\widetilde z_K$ and $\bar Z_1=z_K$.
}

\textbf{Proof of Theorem~\ref{them:ext_reverse}.}
The Koopman-induced degradation distribution defines the joint distribution
\[
    p(z_K,\widetilde z_K)=p(z_K)q_{\mathrm{deg}}(\widetilde z_K|z_{0:K}),
\]
which induces a coupling $\gamma$ between the accurate and degraded latent distributions. Let
\[
    (z_0,z_1)\sim\gamma,
    \qquad z_0=z_K,\quad z_1=\widetilde z_K.
\]
Since both endpoints lie in the same latent vector space, we construct the interpolation
\[
    Z_s=(1-s)z_0+sz_1,
    \qquad s\in[0,1].
\]
For each coupled pair $(z_0,z_1)$, $Z_s$ is continuous in $s$ and satisfies
\[
    Z_0=z_K,
    \qquad Z_1=\widetilde z_K
    \sim q_{\mathrm{deg}}(\cdot|z_{0:K}),
\]
establishing the existence of a continuous stochastic bridge between the accurate and degraded latent states.

Now define its time reversal by
\[
    \bar Z_s:=Z_{1-s},
    \qquad s\in[0,1].
\]
Substituting the interpolation above gives
\[
    \bar Z_s
    =s z_0+(1-s)z_1
    =(1-s)\widetilde z_K+s z_K.
\]
Since $Z_s$ is continuous, $\bar Z_s$ is also continuous, with boundary conditions
\[
    \bar Z_0=\widetilde z_K,
    \qquad\bar Z_1=z_K.
\]
Therefore, $\{\bar Z_s\}_{s\in[0,1]}$ defines a continuous reverse bridge from the Koopman-induced degraded state to the target latent state. This completes the proof.
\hfill $\square$

This theory suggests the existence of a corresponding reverse process for the Koopman-induced degradation distribution, which motivates learning a generative correction model to recover the target latent states.
\section{Dataset Details}
\label{app:equations_datasets}

\paragraph{Navier--Stokes (NS).}
Following~\citet{li2021fno}, we consider the two-dimensional
incompressible Navier--Stokes equations in vorticity form on the
periodic unit torus $\Omega=(0,1)^2$:
\begin{equation}
    \partial_t \omega
    + \mathbf{u}\cdot\nabla\omega
    =
    \nu\Delta\omega + f,
    \qquad
    \nabla\cdot\mathbf{u}=0,
\end{equation}
where $\omega=\nabla\times\mathbf{u}$ denotes the vorticity,
$\mathbf{u}$ is the velocity field, and the viscosity is fixed to
$\nu=10^{-3}$. The external forcing is
\begin{equation}
    f(x_1,x_2)
    =
    0.1\left[
    \sin\left(2\pi(x_1+x_2)\right)
    +
    \cos\left(2\pi(x_1+x_2)\right)
    \right].
\end{equation}
The initial vorticity is sampled from the Gaussian random field
\begin{equation}
    \omega_0
    \sim
    \mathcal{N}\left(
    0,\,
    7^{3/2}(-\Delta+49I)^{-2.5}
    \right),
\end{equation}
with periodic boundary conditions.

We use the publicly released
\texttt{NavierStokes\_V1e-3\_N5000\_T50} dataset
from~\citet{li2021fno}, which contains 5,000 trajectories on a $64\times64$ spatial grid. For all experiments, we use the first 1,000 trajectories, randomly split with seed 777 into 800 training trajectories and 200 test trajectories.

\paragraph{Reaction--diffusion (RD).}
We consider a two-channel periodic reaction--diffusion system on
$\Omega=[0,2\pi)^2$:
\begin{equation}
\begin{aligned}
    \partial_t u &= \Delta u + u - u^3 - v + 0.01,\\
    \partial_t v &= \Delta v + \beta(u-v),
\end{aligned}
\label{eq:appendix_rd}
\end{equation}
where $u$ and $v$ denote the two evolving fields, and $\beta$ controls
their reaction coupling. The initial conditions are sampled from random
truncated Fourier series with periodic boundary conditions.
The dataset contains 200 trajectories on a $64\times64$ spatial grid,
with 40 trajectories for each
$\beta\in\{0.3,0.4,0.5,0.6,0.7\}$. The training file contains 200 trajectories. An independent set of 30 trajectories is used for testing.

\paragraph{2DHeat.}
Following~\citet{koehler2024apebench}, we consider the two-dimensional
heat equation on the periodic unit square $\Omega=[0,1)^2$:
\begin{equation}
    \partial_t h = \nu \Delta h,
    \label{eq:appendix_heat}
\end{equation}
where $h$ denotes the scalar temperature field and $\nu$ is the
trajectory-wise constant diffusivity. No external heat source is applied.
The initial temperature fields are sampled from random Fourier fields with
a cutoff wavenumber of five, zero mean, and unit maximum absolute amplitude,
with periodic boundary conditions.

We use trajectories generated with the diffusion scenario of
APEBench~\citep{koehler2024apebench}, where each trajectory is simulated
on a $64\times64$ periodic spatial grid with a temporal interval of
$\Delta t=0.1$. The dataset contains 250 training trajectories and 60 validation trajectories. Evaluation is performed on 360 independent test trajectories, comprising 150 trajectories at seen diffusivities, 90 at interpolation diffusivities, and 120 at extrapolation diffusivities.
\paragraph{2D electromagnetic scattering (2DEM).}
We consider two-dimensional electromagnetic scattering by dielectric
objects under transverse-magnetic (TM$_z$) polarization, governed by
\begin{equation}
\begin{aligned}
    \frac{\partial E_z}{\partial y}
    &=-\mu_0\frac{\partial H_x}{\partial t},\\
    \frac{\partial E_z}{\partial x}
    &= \mu_0\frac{\partial H_y}{\partial t},\\
    \frac{\partial H_y}{\partial x}
    -\frac{\partial H_x}{\partial y}
    &=\varepsilon_0\varepsilon_r
    \frac{\partial E_z}{\partial t}+J_z,
\end{aligned}
\label{eq:appendix_em2d}
\end{equation}
where $E_z$ is the out-of-plane electric field and $H_x$ and $H_y$
are the in-plane magnetic fields. The medium is lossless and nonmagnetic.
The spatially varying relative permittivity $\varepsilon_r$ determines
the geometry and material property of the dielectric scatterer: the
background is vacuum with $\varepsilon_r=1$, while the scatterer has
relative permittivity selected from $\{6,6.5,7,7.5,8\}$ and one of four
predefined geometries: ellipses, triangles, quadrilaterals, and tri-lobed
structures. Second-order Mur absorbing boundary conditions are imposed
along the four outer boundaries.

The electromagnetic fields are excited by a Gaussian-modulated
sinusoidal point source.

\begin{equation}
    J_z(t)
    =
    -\cos(\omega_c t)
    \exp\left[
    -\frac{4\pi(t-t_0)^2}{\tau^2}
    \right],
\end{equation}
where the center frequency $\omega_c$ is randomly sampled between
$4\times10^{15}$ and $6\times10^{15}\,\mathrm{rad/s}$, the temporal
delay $t_0$ is randomly sampled between $1.8$ and $2.2\,\mathrm{fs}$,
and the pulse width is set to $\tau=t_0/0.8$. The source location is
also varied across trajectories.

The dataset contains $200$ trajectories generated using the FDTD method
on a $128\times128$ Yee grid over an approximately
$2.39\,\mu\mathrm{m}\times2.39\,\mu\mathrm{m}$ domain, with
$\Delta x=\Delta y\approx18.84\,\mathrm{nm}$ and
$\Delta t\approx0.044\,\mathrm{fs}$. We use the
post-excitation source-free evolution of the electromagnetic fields for forecasting.

\section{Evaluation Metrics}
\label{app:metrics}

\paragraph{Normalized Mean Squared Error (NMSE).}
We evaluate forecasting accuracy using the normalized mean squared error
(NMSE) employed in our main comparison table. For each physical-field
channel $c$, let $\mu_c$ and $\sigma_c$ denote the mean and standard
deviation estimated exclusively from the permitted training trajectories
and training-time steps. These statistics are fixed during validation and
testing. The normalized ground-truth and predicted fields are defined as
\begin{equation}
    \widetilde{\mathbf{u}}_{n,t,c}
    =
    \frac{
        \mathbf{u}_{n,t,c}-\mu_c
    }{
        \sigma_c
    },
    \qquad
    \widehat{\widetilde{\mathbf{u}}}_{n,t,c}
    =
    \frac{
        \widehat{\mathbf{u}}_{n,t,c}-\mu_c
    }{
        \sigma_c
    },
    \label{eq:appendix_field_normalization}
\end{equation}
where \(n\), \(t\), and \(c\) index test trajectories, forecast time steps, and physical-field channels.

For a set of forecast time steps $\mathcal{T}$, the reported NMSE is the
mean squared error in this fixed normalized field space:
\begin{equation}
    \mathrm{NMSE}(\mathcal{T})
    =
    \frac{1}{
        N|\mathcal{T}|C|\Omega|
    }
    \sum_{n=1}^{N}
    \sum_{t\in\mathcal{T}}
    \sum_{c=1}^{C}
    \left\|
        \widehat{\widetilde{\mathbf{u}}}_{n,t,c}
        -
        \widetilde{\mathbf{u}}_{n,t,c}
    \right\|_2^2,
    \label{eq:appendix_nmse}
\end{equation}
where $N$ is the number of test trajectories, $C$ is the number of
physical-field channels, and $|\Omega|$ is the number of spatial grid
points. 

\paragraph{Eigenvector-Matrix Condition Number.}
Let $A\in\mathbb{R}^{m\times m}$ denote the learned latent Koopman
generator. Assuming that $A$ is diagonalizable, we write
\begin{equation}
    A = V\Lambda V^{-1},
\end{equation}
where $V=[v_1,\ldots,v_m]$ contains the right eigenvectors of $A$.
Since eigenvectors are defined only up to an arbitrary nonzero scaling,
we normalize each eigenvector to unit Euclidean norm,
\begin{equation}
    v_i \leftarrow \frac{v_i}{\|v_i\|_2}.
\end{equation}
We then compute
\begin{equation}
    \kappa_2(V)
    =
    \|V\|_2\|V^{-1}\|_2
    =
    \frac{\sigma_{\max}(V)}
         {\sigma_{\min}(V)},
\end{equation}
where $\sigma_{\max}(V)$ and $\sigma_{\min}(V)$ denote the largest
and smallest singular values of $V$, respectively.

\section{Training and evaluation details}
\label{app:training_details}
\begin{table*}[t]
\centering
\caption{
Training configurations on NS, RD, 2DHeat, and 2DEM.
}
\small
\renewcommand{\arraystretch}{1.06}
\setlength{\tabcolsep}{3pt}

\begin{tabularx}{\textwidth}{
    @{}
    L{0.13\textwidth}
    L{0.21\textwidth}
    >{\raggedright\arraybackslash}X
    >{\raggedright\arraybackslash}X
    >{\raggedright\arraybackslash}X
    >{\raggedright\arraybackslash}X
    @{}
}
\toprule
\textbf{Category}
& \textbf{Parameter}
& \textbf{NS}
& \textbf{RD}
& \textbf{2DHeat}
& \textbf{2DEM} \\
\midrule

\textbf{\textit{Koopman stage}}
& & & & & \\[-5pt]

& Epochs
& 800
& 800
& 1,000
& 800 \\

& Batch size
& 32
& 32
& 32
& 32 \\

& Optimizer
& Adam
& Adam
& Adam
& Adam \\

& Learning rate
& $1\times10^{-3}$
& $5\times10^{-4}$
& $5\times10^{-4}$
& $5\times10^{-4}$ \\

& Weight decay
& $1\times10^{-4}$
& $1\times10^{-5}$
& $1\times10^{-5}$
& $1\times10^{-5}$ \\

% & LR scheduler
% & StepLR$(25,0.9)$
% & StepLR$(25,0.9)$
% & StepLR$(25,0.9)$
% & \textit{TBD} \\

& Training-time range
& 1--30
& 1--51
& 1--51
& 1--30 \\

& History-state length
& 3
& 3
& 3
& 3 \\

& Prediction horizon
& 12
& 12
& 12
& 12 \\

& $\lambda_{\mathrm{rec}}$
& 1
& 1
& 1
& 1 \\

& $\lambda_{\mathrm{lat}}$
& 0.05
& 0.01
& 0.01
& 0.01 \\

& $\lambda_{\mathrm{pred}}$
& 0.1
& 1
& 1
& 1 \\

\midrule

\textbf{\textit{KoopFlow stage}}
& & & & & \\[-5pt]

& Epochs
& 500
& 500
& 500
& 500 \\

& Batch size
& 64
& 64
& 64
& 64 \\

& Optimizer
& Adam
& Adam
& Adam
& Adam \\

& Learning rate
& $1\times10^{-4}$
& $1\times10^{-4}$
& $1\times10^{-4}$
& $1\times10^{-4}$ \\

& Weight decay
& $1\times10^{-4}$
& $1\times10^{-4}$
& $1\times10^{-4}$
& $1\times10^{-4}$ \\

% & LR scheduler
% & StepLR$(50,0.8)$
% & StepLR$(50,0.8)$
% & StepLR$(50,0.8)$
% & \textit{TBD} \\

\bottomrule
\end{tabularx}

\label{tab:training_configurations}
\end{table*}
\paragraph{Common protocol.}
All KoopFlow experiments use random seed 777 and are conducted on NVIDIA GeForce RTX 4090 GPUs. For all datasets, the first three observed snapshots are used as the historical context, from which the subsequent trajectory is predicted autonomously without access to future ground-truth states. This choice follows the history-length ablation reported by MERLIN~\citep{lu2026merlin}. We use the same three-frame input protocol across all methods to ensure a consistent information budget. Channel-wise normalization statistics are estimated exclusively from the permitted training trajectories and are then fixed for validation and testing. The Koopman stage learns an encoder, a decoder, and a latent evolution operator. In the KoopFlow stage, all Koopman components are frozen, and a generative model is trained to correct the coarse latent trajectory produced by the frozen operator. Detailed dataset-specific training configurations are summarized in Table~\ref{tab:training_configurations}.

\paragraph{Evaluation protocol.}
We evaluate temporal out-of-distribution generalization by separately reporting errors within and beyond the training-time range. For NS and 2DEM the Train and OOD intervals correspond to forecast time steps 4--30 and 31--50, respectively. For RD and 2DHeat, they correspond to time steps 4--51 and 52--101. We further evaluate physical-parameter generalization on 2DHeat using previously unseen diffusivities, including interpolation within the
training parameter range and extrapolation below and above that range. The temporal OOD and physical-parameter generalization results are reported
in Tables~\ref{tab:temporal_ood} and~\ref{tab:heat2d_generalization}.

\section{Benchmark protocol}
\label{app:benchmark_protocol}

\paragraph{Baseline training.}
All baselines use the same trajectory split, visible training-time range, and training-set normalization statistics as KoopFlow. To avoid repetitive descriptions, we detail the training procedure on NS below. The same method-specific objectives and architectures are used on RD, 2DHeat, and 2DEM.

\paragraph{FNO.}
FNO~\citep{li2021fno} uses four Fourier layers with width 32 and 16 Fourier modes.  Each sequence maps three historical observations to the subsequent $T$ time steps. The model is
trained by minimizing the MSE over the complete predicted field block. We train for 500 epochs using Adam with batch size 32, learning rate $10^{-3}$, weight decay $10^{-4}$, gradient clipping at 1.0, and StepLR$(25,0.9)$.

\paragraph{DeepONet.}
DeepONet~\citep{lu2021deeponet} uses 128 basis functions and a three-layer trunk network with hidden width 128. It is trained on the same sequences as FNO using the full-block field MSE.
The optimizer, batch size, learning rate, weight decay, training budget, and learning-rate schedule are also identical to those used for FNO.

\paragraph{KNO.}
KNO~\citep{xiong2024kno} uses an observable width of 32, retains 16 Fourier modes, and applies a linear Koopman operator with power 8. It uses the same
NS training sequences and optimization schedule as FNO. Its objective is
\begin{equation}
    \mathcal{L}_{\mathrm{KNO}}
    =
    5\mathcal{L}_{\mathrm{pred}}
    +
    0.5\mathcal{L}_{\mathrm{rec}},
\end{equation}

\paragraph{MERLIN.}
MERLIN~\citep{lu2026merlin} is trained using length-10 sequences contained entirely within NS time steps 1--30. Its Phase--I stage learns the encoder, decoder, and latent dynamics, while the residual-memory component models
unresolved latent dynamics. Training uses Adam for 500 epochs with an initial learning rate of $5\times10^{-3}$, weight decay $10^{-4}$, and
StepLR$(50,0.8)$.

\paragraph{LoLA.}
LoLA~\citep{rozet2025lola} first learns a field autoencoder and then trains a latent diffusion model on temporal latent bundles. The autoencoder is trained for 500 epochs with 128 updates per epoch, batch size 64, and learning rate $10^{-4}$. The latent diffusion transformer has width 384, depth 8, and temporal bundle length 4. It is trained for 1,024 epochs with
64 updates per epoch, batch size 256, and learning rate $10^{-4}$. 

\paragraph{CFO.}
CFO~\citep{hou2026cfo} uses the official continuous-flow objective and
UNet2D architecture. The 800 NS training trajectories over time steps 1--30
are represented using quintic temporal splines. Training uses 5,000 updates,
batch size 8, learning rate $2\times10^{-4}$, and regularization coefficient
$\gamma=10^{-5}$. Validation is performed every 1,000 updates.

\paragraph{InfoKoop.}
InfoKoop~\citep{cheng2026information} is trained on fixed-length temporal sequences sampled exclusively from the permitted training-time range of each dataset. The sequence length and sampling density are adjusted according to the available training horizon. It is trained for 500 epochs using Adam with batch size 16, learning rate $10^{-3}$, and
StepLR$(20,0.8)$. The training objective is
\begin{equation}
    \mathcal{L}_{\mathrm{InfoKoop}}
    =
    0.1\mathcal{L}_{\mathrm{rec}}
    +
    2\mathcal{L}_{\mathrm{InfoNCE}}
    +
    0.1\mathcal{L}_{\mathrm{ent}},
\end{equation}
with InfoNCE neighborhood size 3, temperature 0.07, and ridge coefficient
0.1.

\subsection{Additional Experimental Results and Analysis}
\label{app:generalization}

\textbf{Analysis of Dissipative Dynamics and Late-Time Prediction on 2DHeat.}
We further analyze the particularly low temporal-OOD error observed on 2DHeat. 2DHeat is governed by a purely dissipative diffusion process without an external heat source. As the system evolves, high-frequency spatial modes are progressively attenuated, and the temperature field becomes increasingly smooth and approaches an equilibrium state. Consequently, the intrinsic complexity of the late-time dynamics decreases, making states at longer horizons easier to predict. Nevertheless, all methods are evaluated under the same normalization and temporal split, and KoopFlow consistently yields the lowest forecasting error in this late-time regime, indicating that it accurately captures the dissipative evolution without substantial accumulation of long-horizon error.

\begin{figure}[h]
\begin{center}
    \includegraphics[width=1\linewidth]{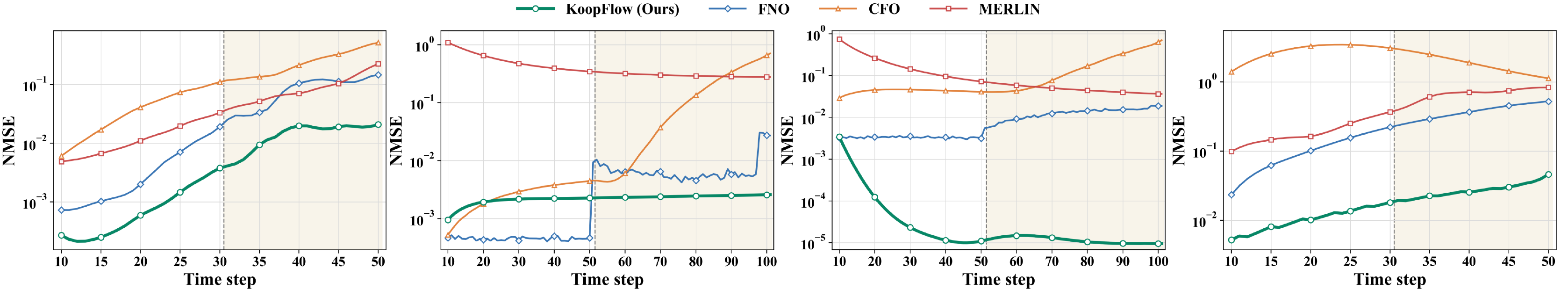}
\end{center}
    \caption{Error comparison of KoopFlow with strong baselines on NS, RD, 2DHeat, and 2DEM.}
    \label{fig:ns_accumated_error_all}
\end{figure}
\textbf{Long-horizon rollout stability.}
Figure~\ref{fig:ns_accumated_error_all} further compares the temporal evolution of forecasting errors across multiple methods. Beyond the training-time range, KoopFlow exhibits substantially more stable error dynamics than the competing methods. On NS, its error increases after entering the temporal-OOD regime but gradually saturates, rather than continuing to accumulate over the rollout horizon. On RD, KoopFlow maintains an almost constant error level across the training/OOD boundary, while several baselines exhibit abrupt error growth or pronounced fluctuations at later time steps. The stabilization effect is even more evident on 2DHeat, where the KoopFlow error rapidly decreases to a low level and remains nearly unchanged throughout the remaining rollout. In contrast, the errors of several baselines continue to grow with the prediction horizon. On 2DEM, although the error of the CFO method shows a descending trend in the later stages, its NMSE exceeds 1, indicating substantial prediction errors. In contrast, the error of KoopFlow remains consistently much lower than other comparative methods. These results suggest that KoopFlow not only improves pointwise forecasting accuracy but also suppresses progressive error accumulation, leading to more stable long-horizon rollouts under temporal distribution shift.
\raggedbottom
\subsection{More Qualitative Examples}
\label{app:more_qualitative_examples}
This section presents additional qualitative examples comparing KoopFlow against the baseline models, providing a visual comparison of their predictive performance across different dynamical systems, including NS, RD, 2DHeat and 2DEM. These examples further illustrate the differences in prediction fidelity and the ability to capture spatiotemporal dynamics over extended prediction horizons.

\begin{figure}[hp]
\begin{center}
    \includegraphics[width=1\linewidth]{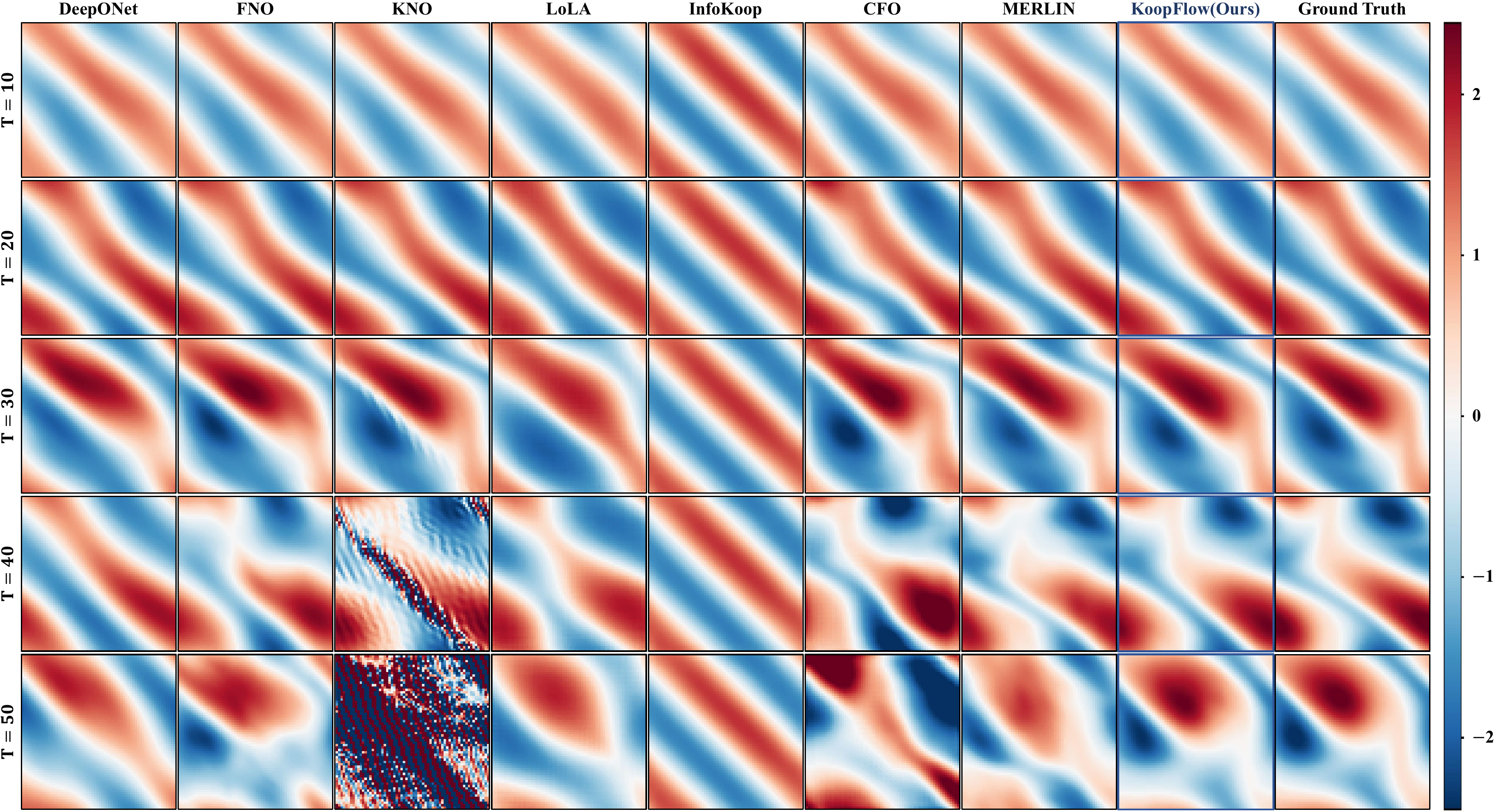}
\end{center}
    \caption{Qualitative comparison of KoopFlow with other baselines on NS.}
    \label{fig:ns_traj451_final_time_field_error_all}
\end{figure}

\begin{figure}[hp]
\begin{center}
    \includegraphics[width=1\linewidth]{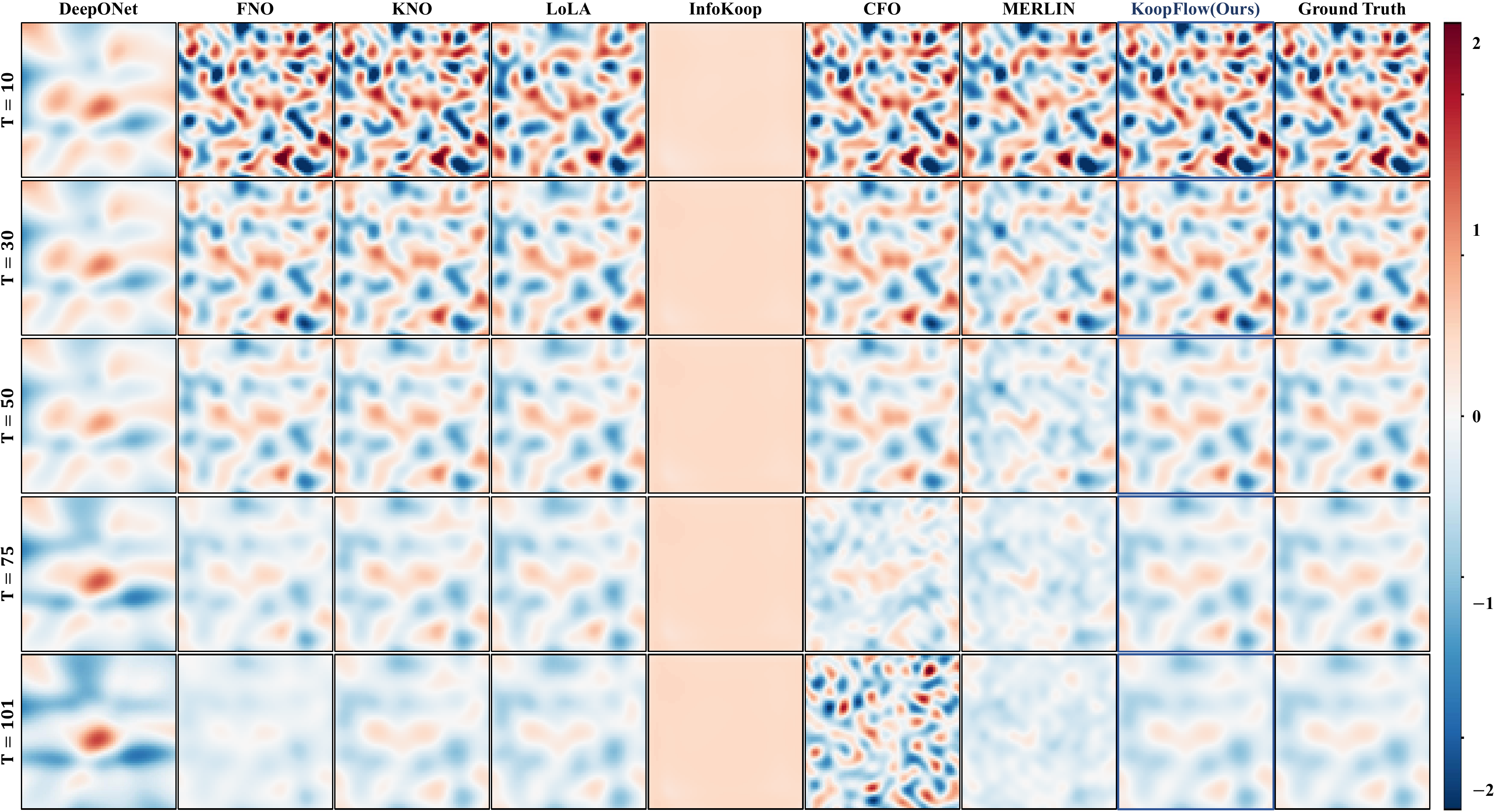}
\end{center}
    \caption{Qualitative comparison of KoopFlow with other baselines on RD.}
    \label{fig:rd_traj451_final_time_field_error}
\end{figure}

\begin{figure}[hp]
\begin{center}
    \includegraphics[width=1\linewidth]{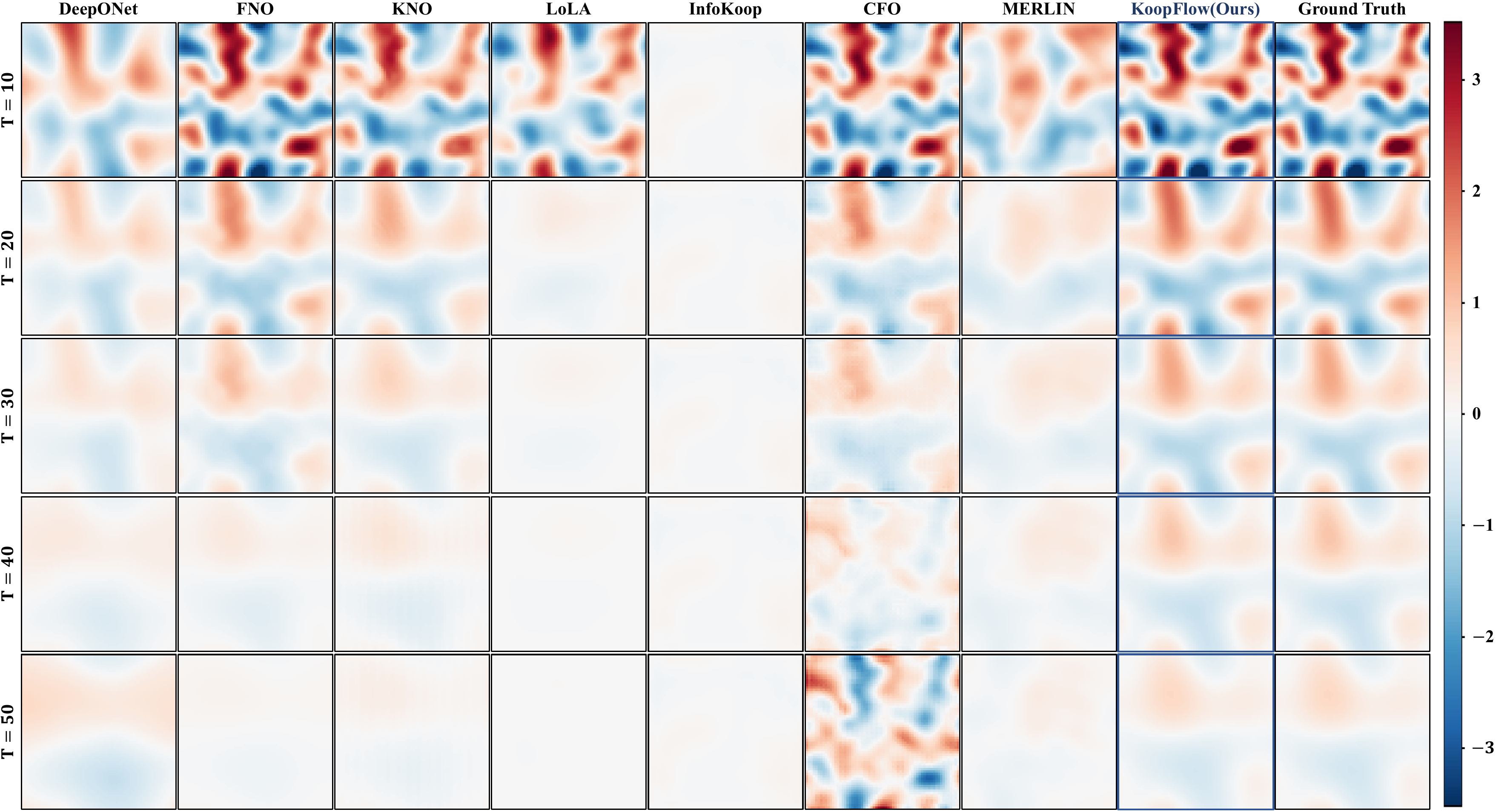}
\end{center}
    \caption{Qualitative comparison of KoopFlow with other baselines on 2DHeat.}
    \label{fig:heat_final_time_field_error_all}
\end{figure}

\begin{figure}[hp]
\begin{center}
    \includegraphics[width=1\linewidth]{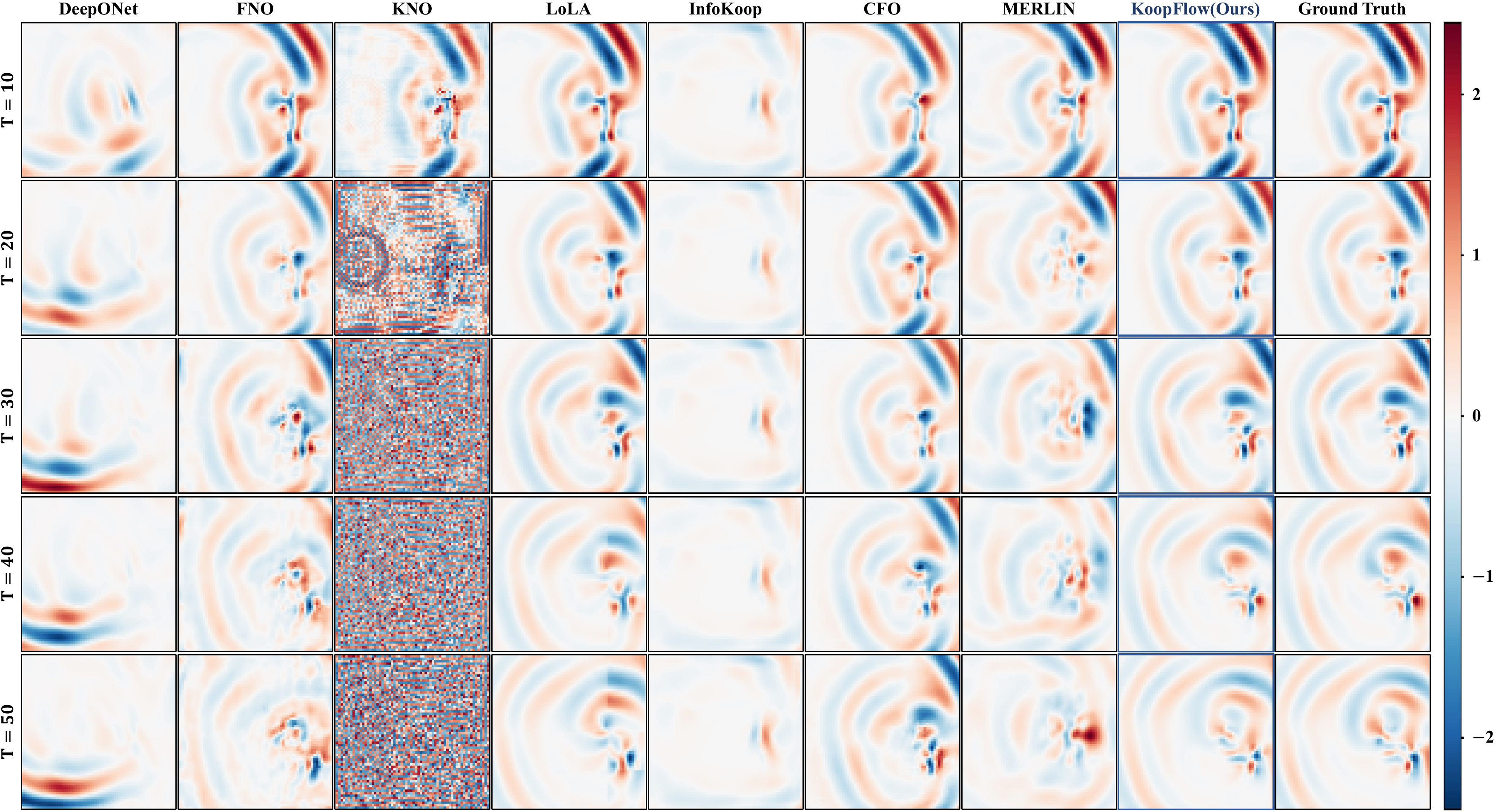}
\end{center}
    \caption{Qualitative comparison of KoopFlow with other baselines on 2DEM.}
    \label{fig:ez_final_time_field_error_all}
\end{figure}
%%%%%%%%%%%%%%%%%%%%%%%%%%%%%%%%%%%%%%%%%%%%%%%%%%%%%%%%%%%%%%%%%%%%%%%%%%%%%%%
%%%%%%%%%%%%%%%%%%%%%%%%%%%%%%%%%%%%%%%%%%%%%%%%%%%%%%%%%%%%%%%%%%%%%%%%%%%%%%%

\end{document}